\pdfoutput=1

\documentclass[aps,showpacs,twocolumn,prd,superscriptaddress,nofootinbib]{revtex4-1}

\usepackage{amsmath}
\usepackage{amsfonts}
\usepackage{amssymb}
\usepackage{tabularx}
\usepackage{booktabs}
\usepackage{hyperref}
\usepackage{xcolor}
\usepackage[normalem]{ulem}
\usepackage{graphicx}
\makeatletter
\newcommand{\bal}{\@ifstar{\@bals}{\@bal}}
\def\@bals#1\eal{\begin{align*}#1\end{align*}}
\def\@bal#1\eal{\begin{align}#1\end{align}}
\makeatletter
\newcommand{\del}{\nabla}
\newcommand{\enDash}{\mbox{--}}

\newcommand{\Order}{\mathcal{O}}

\begin{document}

\newcolumntype{L}[1]{>{\raggedright\arraybackslash}p{#1}}
\newcolumntype{C}[1]{>{\centering\arraybackslash}p{#1}}
\newcolumntype{R}[1]{>{\raggedleft\arraybackslash}p{#1}}

\title{Distinguishing Boson Stars from Black Holes and Neutron Stars from Tidal Interactions in Inspiraling Binary Systems}
\author{Noah Sennett}
\affiliation{Max Planck Institute for Gravitational Physics (Albert Einstein Institute), Am M\"uhlenberg 1, Potsdam-Golm, 14476, Germany}
\affiliation{Department of Physics, University of Maryland, College Park, Maryland 20742, USA}
\author{Tanja Hinderer}
\affiliation{Max Planck Institute for Gravitational Physics (Albert Einstein Institute), Am M\"uhlenberg 1, Potsdam-Golm, 14476, Germany}
\author{Jan Steinhoff}
\affiliation{Max Planck Institute for Gravitational Physics (Albert Einstein Institute), Am M\"uhlenberg 1, Potsdam-Golm, 14476, Germany}
\author{Alessandra Buonanno}
\affiliation{Max Planck Institute for Gravitational Physics (Albert Einstein Institute), Am M\"uhlenberg 1, Potsdam-Golm, 14476, Germany}
\affiliation{Department of Physics, University of Maryland, College Park, Maryland 20742, USA}
\author{Serguei Ossokine}
\affiliation{Max Planck Institute for Gravitational Physics (Albert Einstein Institute), Am M\"uhlenberg 1, Potsdam-Golm, 14476, Germany}
\email{nsennett@umd.edu}
\date{\today}

\begin{abstract}
Binary systems containing boson stars---self-gravitating configurations of a complex scalar field--- can potentially mimic black holes or neutron stars as gravitational-wave sources. We investigate the extent to which tidal effects in the gravitational-wave signal can be used to discriminate between these standard sources and boson stars. We consider spherically symmetric boson stars within two classes of scalar self-interactions: an effective-field-theoretically motivated quartic potential and a solitonic potential constructed to produce very compact stars. We compute the tidal deformability parameter characterizing the dominant tidal imprint in the gravitational-wave signals for a large span of the parameter space of each boson star model, covering the entire space in the quartic case, and an extensive portion of interest in the solitonic case. We find that the tidal deformability for boson stars with a quartic self-interaction is bounded below by $\Lambda_\text{min}\approx 280$ and for those with a solitonic interaction by $\Lambda_\text{min}\approx 1.3$. We summarize our results as ready-to-use fits for practical applications. Employing a Fisher matrix analysis, we estimate the precision with which Advanced LIGO and third-generation detectors can measure these tidal parameters using the inspiral portion of the signal. We discuss a novel strategy to improve the distinguishability between black holes/neutrons stars and boson stars by combining tidal deformability measurements of each compact object in a binary system, thereby eliminating the scaling ambiguities in each boson star model. Our analysis shows that current-generation detectors can potentially distinguish boson stars with quartic potentials from black holes, as well as from neutron-star binaries if they have either a large total mass or a large (asymmetric) mass ratio. Discriminating solitonic boson stars from black holes using only tidal effects during the inspiral will be difficult with Advanced LIGO, but third-generation detectors should be able to distinguish between binary black holes and these binary boson stars. 
\end{abstract}

\maketitle

\section{Introduction}

Observations of gravitational waves (GWs) by Advanced LIGO \cite{TheLIGOScientific:2014jea}, soon to be joined by Advanced Virgo \cite{TheVirgo:2014hva}, KAGRA \cite{Aso:2013eba}, and LIGO-India \cite{LIGOIndia}, open a new window to the strong-field regime of general relativity (GR). A major target for these detectors are the GW signals produced by the coalescences of binary systems of compact bodies. Within the standard astrophysical catalog, only black holes (BHs) and neutron stars (NSs) are sufficiently compact to generate GWs detectable by current-generation ground-based instruments. To test the dynamical, non-linear regime of gravity with GWs, one compares the relative likelihood that an observed signal was produced by the coalescence of BHs or NSs as predicted by GR against the possibility that it was produced by the merger of either: (a) BHs or NSs in alternative theories of gravity or (b) exotic compact objects in GR. In this paper, we pursue tests within the second class. Several possible exotic objects have been proposed that could mimic BHs or NSs, including boson stars (BSs)~\cite{Schunck:2003kk,Liebling:2012fv}, gravastars~\cite{Mazur:2001fv,Mazur:2004fk}, quark stars~\cite{Itoh:1970uw}, and axion stars~\cite{Eby:2017xaw,Iwazaki:1999vi}.

The coalescence of a binary system can be classified into three phases--- the inspiral, merger, and ringdown--- each of which can be modeled with different tools. The inspiral describes the early evolution of the binary and can be studied within the post-Newtonian (PN) approximation, a series expansion in powers of the relative velocity $v/c$ (see Ref.~\cite{Blanchet:2013haa} and references within). As the binary shrinks and eventually merges, strong, highly-dynamical gravitational fields are generated; the merger is only directly computable using numerical relativity (NR). Finally, during ringdown, the resultant object relaxes to an equilibrium state through the emission of GWs whose (complex) frequencies are given by the object's quasinormal modes (QNMs), calculable through perturbation theory (see Ref.~\cite{Kokkotas:1999bd} and references within). Complete GW signals are built by synthesizing results from these three regimes from first principles with the effective-one-body (EOB) formalism~\cite{Buonanno:1998gg,Bohe:2016gbl} or phenomenologically, through frequency-domain fits~\cite{Ajith:2007kx,Khan:2015jqa} of inspiral, merger and ringdown waveforms.

An understanding of how exotic objects behave during each of these phases is necessary to determine whether GW detectors can distinguish them from conventional sources (i.e., BHs or NSs). Significant work in this direction has already been completed. The structure of spherically-symmetric compact objects is imprinted in the PN inspiral through tidal interactions that arise at 5PN order (i.e., as a $(v/c)^{10}$ order correction to the Newtonian dynamics). Tidal interactions are characterized by the object's tidal deformability, which has recently been computed for gravastars \cite{Pani:2015tga,Uchikata:2016qku} and ``mini'' BSs~\cite{Mendes:2016vdr}. During the completion of this work, an independent investigation on the tidal deformability of several classes of exotic compact objects, including examples of the BS models considered here, was performed in Ref.~\cite{Cardoso:2017cfl}; details of the similarities and differences to this work are discussed in Sec.~\ref{sec:Conclusions} below. Additional signatures of exotic objects include the magnitude of the spin-induced quadrupole moment and the absence of tidal heating. The possibility of discriminating BHs from exotic objects with these two effects was discussed in Refs.~\cite{Krishnendu:2017shb} and~\cite{Maselli:2017cmm}, respectively---we will not consider these effects in this paper. The merger of BSs has been studied using NR in head-on collisions \cite{Palenzuela:2006wp,Cardoso:2016oxy} and following circular orbits \cite{Palenzuela:2007dm}. The QNMs have been computed for BSs~\cite{Yoshida:1994xi,Hawley:2000dt,Macedo:2013jja} and gravastars~\cite{Chirenti:2007mk,Pani:2009ss,Chirenti:2016hzd}. 

In this paper, we compute the tidal deformability of two models of BSs: 
 ``massive'' BSs \cite{Colpi:1986ye} characterized by a quartic self-interaction and non-topological solitonic BSs \cite{Friedberg:1986tq}. The self-interactions investigated here allow for the formation of compact BSs, in contrast to the ``mini'' BSs considered in Ref.~\cite{Mendes:2016vdr}. We perform an extensive analysis of the BS parameter space within these models, thereby going beyond previous work in Ref.~\cite{Cardoso:2017cfl}, which was limited to a specific choice of parameter characterizing the self-interaction for each model. Special consideration must be given to the choice of the numerical method because BSs are constructed by solving stiff differential equations---we employ relaxation methods to overcome this problem~\cite{press1992numerical}. Our new findings show that for massive BSs, the tidal deformability $\Lambda$ (defined below) is bounded below by $\Lambda_\text{min}\approx 280$ for stable configurations, while for solitonic BSs the deformability can reach $\Lambda_\text{min}\approx 1.3$. For comparison, the deformability of NSs is $\Lambda_\text{NS}\gtrsim \mathcal{O}(10)$ and for BHs $\Lambda_\text{BH}=0$. We compactly summarize our results as fits for convenient use in future gravitational wave data analysis studies. In addition, we employ the Fisher matrix formalism to study the prospects for distinguishing BSs from NSs or BHs with current and future gravitational-wave detectors based on tidal effects during the inspiral. Prospective constraints on the combined tidal deformability parameters of both objects in a binary were also shown for two fiducial cases in Ref.~\cite{Cardoso:2017cfl}. Our findings are consistent with the conclusions drawn in Ref.~\cite{Cardoso:2017cfl}; we discuss a new type of analysis that can strengthen the claims made therein on the distinguishability of BSs from BHs and NSs by combining information on each body in a binary system.
 
The paper is organized as follows. Section~\ref{sec:BSbasics} introduces the BS models investigated herein. We provide the necessary formalism for computing the tidal deformability in Sec.~\ref{sec:BackgroundandPerturbation}, and describe the numerical methods we employ in Sec.~\ref{sec:NumericalMethods}. In Sec.~\ref{sec:Results}, we compute the tidal deformability, providing results that range from the weak-coupling limit to the strong-coupling limit as well as numerical fits for the tidal deformability. Finally, in Sec.~\ref{sec:Constraints} we discuss the prospects of testing the existence of stellar-mass BSs using GW detectors and provide some concluding remarks in Sec.~\ref{sec:Conclusions}.

We use the signature $(-,+,+,+)$ for the metric and natural units $\hbar=G=c=1$, but explicitly restore factors of the Planck mass  $m_\text{Planck}=\sqrt{\hbar c/G}$ in places to improve clarity.
The convention for the curvature tensor is such that 
$\nabla_\beta \nabla_\alpha a_\mu - \nabla_\beta \nabla_\beta a_\mu = R^\nu{}_{\mu\alpha\beta} a_{\nu}$, where $\nabla_\alpha$ is the covariant derivative and $a_\mu$ is a generic covector. 

\section{Boson star basics}\label{sec:BSbasics}
Boson stars---self-gravitating configurations of a (classical) complex scalar field---have been studied extensively in the literature, both as potential dark matter candidates and as tractable toy models for testing generic properties of compact objects in GR. Boson stars are described by the Einstein-Klein-Gordon action
\begin{align}
S=\int d^4 x \sqrt{-g}\left[\frac{R}{16 \pi}-\del^\alpha \Phi \nabla_\alpha \Phi^*-V(|\Phi|^2)\right] ,\label{eq:Action}
\end{align}
where ${}^*$ denotes complex conjugation. The only experimentally confirmed elementary scalar field is the Higgs boson \cite{Chatrchyan:2012xdj,Aad:2012tfa}, which is an unlikely candidate to form a BS because it readily decays to $W$ and $Z$ bosons. However, other massive scalar fields have been postulated in many theories beyond the Standard Model, e.g., bosonic superpartners predicted by supersymmetric extensions \cite{Roszkowski:1993by}.

The Einstein equations derived from the action~\eqref{eq:Action} are given by
\begin{align}
R_{\alpha \beta}-\frac{1}{2} g_{\alpha \beta} R= 8\pi T^\Phi_{\alpha \beta},\label{eq:EFEbackground}
\end{align}
with
\begin{align}
\begin{split}
T^\Phi_{\alpha \beta}=&\del_\alpha \Phi^* \del_\beta \Phi+\del_\beta \Phi^*\del_\alpha \Phi\\
&-g_{\alpha \beta}(\nabla^\gamma \Phi^* \nabla_\gamma \Phi+V(|\Phi|^2)).\label{eq:TPhidef}
\end{split}
\end{align}
The accompanying Klein-Gordon equation is
\begin{align}
\nabla^\alpha \nabla_\alpha \Phi=\frac{d V}{d |\Phi |^2}\Phi, \label{eq:KGbackground}
\end{align}
along with its complex conjugate.

The earliest proposals for a BS contained a single non-interacting scalar field~\cite{Kaup:1968zz,Feinblum:1968nwc,Ruffini:1969qy}, that is
\begin{align}
V\left(|\Phi |^2\right)= \mu^2 |\Phi |^2,\label{eq:MiniV}
\end{align}
where $\mu$ is the mass of the boson. The free Einstein-Klein-Gordon action also describes the second-quantized theory of a real scalar field; thus, this class of BS can also be interpreted as a gravitationally bound Bose-Einstein condensate~\cite{Ruffini:1969qy}. The maximum mass for BSs with the potential given in Eq.~\eqref{eq:MiniV} is ${M_\text{max}\approx 0.633 m_\text{Planck}^2/\mu}$,  or in units of solar mass, ${M_\text{max} / M_\odot \approx 85 \text{peV} / \mu}$. The corresponding compactness for this BS is ${C_\text{max}\approx 0.08}$ \cite{Kaup:1968zz}.\footnote{Formally, BSs have no surface, so the notion of a radius (and hence compactness) is inherently ambiguous. One common convention is to define the radius as that of a shell containing a fixed fraction of the total mass of the star (e.g., $R_{99}$ where ${m(r=R_{99})=0.99\,m(r=\infty)}$). To avoid this ambiguity, our results are given in terms of quantities that can be extracted directly from the asymptotic geometry of the BS: the total mass $M$ and dimensionless tidal deformability $\Lambda$ (defined below).} Because this maximum mass scales more slowly with $\mu$ than the Chandrasekhar limit for a degenerate fermionic star ${M_\text{CH}\sim m_\text{Planck}^3/m_{\rm Fermion}^2}$, this class of BSs is referred to as \textit{mini BSs}. The tidal deformability was computed in this model in Ref.\ \cite{Mendes:2016vdr}

Since the seminal work of the 1960s~\cite{Kaup:1968zz,Feinblum:1968nwc,Ruffini:1969qy}, BSs with various scalar self-interactions have been studied. We consider two such models in this paper, both which reduce to mini BSs in the weak-coupling limit. The first BS model we consider is \textit{massive BSs}~\cite{Colpi:1986ye}, with a potential given by

\begin{align}
V_{\rm massive}(|\Phi|^2)=\mu^2 |\Phi|^2+\frac{\lambda}{2}|\Phi|^4,\label{eq:MassiveV}
\end{align}
which is repulsive for $\lambda\ge 0$. In the strong-coupling limit $\lambda \gg \mu^2 / m_\text{Planck}^2$, spherically symmetric BSs obtain a maximum mass of ${M_\text{max}\approx0.044\sqrt{\lambda}m_\text{Planck}^3/\mu^2}$~\cite{Colpi:1986ye}. In units of the solar mass $M_\odot$ this reads ${M_\text{max} / M_\odot \approx \sqrt{\lambda} (0.3 \text{GeV} / \mu)^2}$. Such configurations are roughly as compact as NSs, with an effective compactness of ${C_\text{max}\approx 0.158}$ \cite{Guzman:2005bs,AmaroSeoane:2010qx}. This BS model is a natural candidate from an effective-field-theoretical perspective because the potential in Eq.~\eqref{eq:MassiveV} contains all renormalizable self-interactions for a scalar field, i.e., other interactions that scale as higher powers of $|\Phi|$ are expected to be suppressed far from the Planck scale. The ``natural'' values of $\lambda \sim 1$ and $\mu \ll m_\text{Planck}$ yield the strong-coupling limit of the potential ~\eqref{eq:MassiveV}. Because it is the most theoretically plausible BS model, we investigate the strong-coupling regime of this interaction in detail in Section~\ref{sec:MassiveResults}.

The second class of BS that we consider is the \textit{solitonic BS} model \cite{Friedberg:1986tq}, characterized by the potential
\begin{align}
V_{\rm solitonic}(|\Phi|^2)=\mu^2 |\Phi|^2\left(1-\frac{2 |\Phi|^2}{\sigma_0^2}\right)^2 .\label{eq:SolitonicV}
\end{align}
This potential admits a false vacuum solution at ${|\Phi |=\sigma_0/\sqrt{2}}$. One can construct spherically symmetric BSs whose interior closely resembles this false vacuum state and whose exterior is nearly vacuum ${|\Phi |\approx 0}$; the transition between the false vacuum and true vacuum occurs over a surface of width ${\Delta r\sim \mu^{-1}}$.
In the strong-coupling limit $\sigma_0 \ll m_\text{Planck}$, the maximum mass
of non-rotating BSs is $M_\text{max}\approx 0.0198 m_\text{Planck}^4/(\mu\sigma_0^2)$, or
$M_\text{max} / M_\odot \approx (\mu / \sigma_0)^2 (0.7 \text{PeV} / \mu)^3$~\cite{Friedberg:1986tq}. The corresponding compactness $C_\text{max}\approx 0.349$ approaches that of a BH
$C_\text{BH}=1/2$~\cite{Friedberg:1986tq}.\footnote{This compactness is still lower than the theoretical Buchdahl limit of $C \leq 4/9$ for isotropic perfect fluid stars that respect the strong energy condition \cite{Buchdahl:1959zz}.}
The main motivation for considering the potential (\ref{eq:SolitonicV}) is as a model of
very compact objects that could even possess a light-ring when $C>1/3$. In this paper, we will only consider solitonic BSs as potential BH mimickers, as NSs could be mimicked by the more natural massive BS model.

In this paper, we restrict our attention to only non-rotating BSs. Axisymmetric (rotating) BSs have been constructed for the models we consider~\cite{Yoshida:1997qf,Ryan:1996nk,Kleihaus:2005me,Grandclement:2014msa}, but these solutions are significantly more complex than those that are spherically symmetric (non-rotating). The energy density of a rotating BS forms a toroidal topology, vanishing at the star's center. Because its angular momentum is quantized, a rotating BS cannot be constructed in the slow-rotation limit, i.e. by adding infinitesimal rotation to a spherically symmetric solution~\cite{Kobayashi:1994qi}.

\section{Tidal Perturbations of Spherically-symmetric boson stars}\label{sec:BackgroundandPerturbation}
We consider linear tidal perturbations of a non-rotating BSs.  We work within the adiabatic limit, that is we assume that the external tidal field varies on timescales much longer than any oscillation period of the star or relaxation timescale to reach a microphysical equilibrium. These conditions are typically satisfied during the inspiral of compact binaries. Close to merger, the assumptions concerning the separation of timescales can break down and the tides can become
dynamical~\cite{Hinderer:2016eia, Steinhoff:2016rfi, Kokkotas:1995xe, Lai:1993di}; we ignore these complications here.
The computation of the tidal deformability of NSs in general relativity was first addressed in
Refs.~\cite{Hinderer:2007mb, Flanagan:2007ix} and was extended in Refs.\ \cite{Damour:2009vw, Binnington:2009bb}.

\subsection{Background configuration}
Here we review the equations describing a spherically symmetric BS \cite{Kaup:1968zz, Colpi:1986ye, Schunck:2003kk},
which is the background configuration that we use to compute the tidal perturbations in the following subsection.
We follow the presentation in Ref.\ \cite{Macedo:2013jja}.
The metric written in polar-areal coordinates reads
\begin{align}
ds_0^2=-e^{v(r)} dt^2+e^{u(r)} dr^2+r^2 (d\theta^2+\sin^2\theta d\varphi^2). \label{eq:BackgroundMetric}
\end{align}
As an ansatz for the background scalar field, we use the decomposition
\begin{align}
\Phi_0(t,r)=\phi_0(r)e^{-i \omega t}.  \label{eq:BackgroundPhi}
\end{align}
Inserting Eqs.~\eqref{eq:BackgroundMetric} and~\eqref{eq:BackgroundPhi} into Eqs.~\eqref{eq:EFEbackground}--\eqref{eq:KGbackground} gives
\begin{subequations}
\begin{gather}
e^{-u}\left(-\frac{u'}{r}+\frac{1}{r^2}\right)-\frac{1}{r^2}=-8 \pi \rho,\label{eq:TOVu}\\
e^{-u}\left(\frac{v'}{r}+\frac{1}{r^2}\right)-\frac{1}{r^2}=8\pi p_\text{rad},\label{eq:TOVv}\\
\phi''_0+\left(\frac{2}{r}+\frac{v'-u'}{2}\right)\phi'_0=e^{u}\left(U_0-\omega^2e^{-v}\right)\phi_0,\label{eq:TOVphi}
\end{gather}
\end{subequations}
where a prime denotes differentiation with respect to $r$, $U_0=U(\phi_0)$, $U(\phi)=dV/d|\Phi|^2$.
Because the coefficients in Eq.~(\ref{eq:TOVphi}) are real numbers, we can restrict
$\phi_0(r)$ to be a real function without loss of generality. We have also defined the effective density and pressures
\begin{align}
\rho\equiv& -{T^{\Phi}}_t^t=\omega^2 e^{-v}\phi_0^2+e^{-u}(\phi_0')^2+V_0,\label{eq:DensityDef}\\
p_\text{rad}\equiv& {T^{\Phi}}_r^r=\omega^2 e^{-v}\phi_0^2+e^{-u}(\phi_0')^2-V_0,\\
p_\text{tan}\equiv& {T^{\Phi}}_\theta^\theta=\omega^2 e^{-v}\phi_0^2-e^{-u}(\phi_0')^2-V_0,
\end{align}
where $V_0=V(\phi_0)$. Note that BSs behave as anisotropic fluid stars with pressure anisotropy given by
\begin{align}
\Sigma=p_\text{rad}-p_\text{tan}=2 e^{-u}(\phi_0')^2.\label{eq:Anisotropy}
\end{align}
An additional relation derived from Eqs.~\eqref{eq:EFEbackground}--\eqref{eq:BackgroundPhi} that will be used to simplify the perturbation equations discussed in the next subsection is
\begin{align}
p'_\text{rad}=-\frac{(p_\text{rad}+\rho)}{2r}\left[e^u\left(1+8\pi r^2 p_\text{rad}\right)-1\right]-\frac{2\Sigma}{r}. \label{eq:TOVprad}
\end{align}

We restrict our attention to ground-state configurations of the BS, in which $\phi_0(r)$ has no nodes. The background fields exhibit the following asymptotic behavior
\begin{subequations}\label{eq:BackAsymptotics}
\begin{align}
\lim_{r\rightarrow 0} m(r)&\sim r^3, &&\lim_{r\rightarrow \infty} m(r)\sim M,\label{eq:uAsymptotics}\\
\lim_{r\rightarrow 0} v(r)&\sim v^{(c)}, &&\lim_{r\rightarrow \infty} v(r) \sim 0,\label{eq:vAsymptotics}\\
\lim_{r\rightarrow 0} \phi_0(r)&\sim \phi_0^{(c)}, &&\lim_{r\rightarrow \infty} \phi_0(r)\sim \frac{1}{r} e^{-r\sqrt{\mu^2-\omega^2} },\label{eq:Phi0Asymptotics}
\end{align}
\end{subequations}
where $M$ is the BS mass, $v^{(c)}$ and $\phi_0^{(c)}$ are constants, and $m(r)$ is defined such that
\begin{align}
e^{-u(r)}=\left(1-\frac{2 m(r)}{r}\right).
\end{align}

\subsection{Tidal perturbations}
We now consider small perturbations to the metric and scalar field defined such that
\begin{align}
g_{\alpha \beta}&=g_{\alpha \beta}^{(0)}+h_{\alpha \beta}, \label{eq:gpert}\\
\Phi&=\Phi_0+\delta \Phi. \label{eq:phipert}
\end{align}
We restrict our attention to static perturbations in the polar sector, which describe the effect of an external electric-type tidal field. Working in the Regge-Wheeler gauge \cite{Regge:1957td}, the perturbations take the form
\begin{subequations}
\label{eq:hansatz}
\begin{align}
\begin{split}
h_{\alpha \beta}dx^\alpha dx^\beta=&\sum_{l\ge |m|}Y_{lm}(\theta,\varphi)\left[e^{v} h_0(r)  dt^2
\right.\\
&\left.+e^{u} h_2(r) dr^2 +r^2 k(r)(d\theta^2+r \sin^2 \theta d\varphi^2)\right],
\end{split}
\end{align}
and
\begin{align}
\delta \Phi=&\sum_{l\ge |m|} \frac{\phi_1(r)}{r}Y_{l m}(\theta,\varphi) e^{- i \omega t}, 
\end{align}
\end{subequations}
where $Y_{lm}$ are scalar spherical harmonics.

We insert the perturbed metric and scalar field from Eqs.~\eqref{eq:gpert}--\eqref{eq:hansatz} into the Einstein and Klein-Gordon equations, Eqs.~\eqref{eq:EFEbackground} and~\eqref{eq:KGbackground}, and expand to first order in the perturbations. For the metric functions, the ${(\theta,\phi)\text{-component}}$ of the Einstein equations gives $h_2=h_0$, and the $(r,r)$- and ${(r, \theta)\text{-components}}$ can be used to algebraically eliminate $k$ and $k^\prime$ in favor of $h_0$ and its derivatives. Finally, the ${(t,t)\text{-component}}$ leads to the following second-order differential equation:
\begin{widetext}
\begin{align}
\begin{split}
&h_0''+\frac{e^{u}h_0'}{r} \left(1+e^{-u}-8 \pi  r^2 
   V_0\right)-\frac{32 \pi  e^{u}\phi _1}{r^2} \left[\phi _0' \left(-1+e^{-u}-8 \pi  r^2
    p_\text{rad}\right)+r \phi
   _0 \left(U_0 -2 \omega ^2e^{-v}\right)\right]\\
&+\frac{h_0 e^u}{r^2}\left[-16 \pi  r^2 V_0-l (l+1)-e^u(1-e^{-u}+8 \pi r^2 p_\text{rad})^2+64 \pi  r^2 \omega ^2 \phi _0^2 e^{-v}\right]=0,\label{eq:Perturbationh}
\end{split}
\end{align}
where we have also used the background equations~\eqref{eq:TOVu},~\eqref{eq:TOVv}, and~\eqref{eq:TOVprad}. From the linear perturbations to the Klein-Gordon equation, together with the results for the metric perturbations and the background equations, we obtain
\begin{align}
\begin{split}
&\phi _1''+\frac{e^u \phi _1'}{r} \left(1-e^{-u}-8 \pi  r^2
   V_0\right)-e^{u}h_0 \left[\phi _0' \left(-1+e^{-u}-8 \pi  r^2 p_\text{rad}\right)+r\phi _0
   \left(U_0 -2 \omega ^2 e^{-v} \right)\right]\\
  & +\frac{e^u\phi _1}{r^2} \left[8 \pi r^2  V_0-1+e^{-u}-l
   (l+1)-r^2 \left(U_0+2 W_0 \phi _0^2\right)+r^2 e^{-v} \omega ^2-32
   \pi  e^{-u}r^2 \left(\phi _0'\right)^2\right]=0,\label{eq:Perturbationphi}
\end{split}\end{align}
\end{widetext}
where $W_0=W(\phi_0)$ with $W(\phi)=dU/d|\Phi|^2$.
These perturbation equations were also independently derived in Ref.~\cite{Cardoso:2017cfl} and are a special case of generic linear perturbations
considered in the context of QNMs (see, e.g., Refs.~\cite{Yoshida:1994xi,Hawley:2000dt,Macedo:2013jja}). As a check, we combined the three first-order and one algebraic constraint for the spacetime perturbations from Ref.\ \cite{Macedo:2013jja} into one second-order equation for $h_0$, which agrees with Eq.~(\ref{eq:Perturbationh}) in the limit of static perturbations. 
For the special case of mini BSs, the tidal perturbation equations were also
obtained in Ref.\ \cite{Mendes:2016vdr}. 

The perturbations exhibit the following asymptotic behavior \cite{Macedo:2013jja}
\begin{subequations}\label{eq:PertAsymptotics}
\begin{align}
\lim_{r\rightarrow 0} h_0(r)&\sim r^l,\\
\lim_{r\rightarrow \infty} h_0(r)&\sim c_1 \left(\frac{r}{M}\right)^{-(l+1)}+c_2 \left(\frac{r}{M}\right)^l,\label{eq:h0Asymptotics}\\
\lim_{r\rightarrow 0} \phi_1(r)&\sim r^{l+1},\\
\lim_{r\rightarrow \infty} \phi_1(r)&\sim r^{M \mu^2/\sqrt{\mu^2-\omega^2}}e^{-r\sqrt{\mu^2-\omega^2} }.
\label{eq:Phi1Asymptotics}
\end{align}
\end{subequations}

\subsection{Extracting the tidal deformability}
The BS tidal deformability can be obtained in a similar manner as with NSs
\cite{Hinderer:2007mb, Damour:2009vw, Binnington:2009bb}. Working in the (nearly) vacuum region far from the center of the BS, the formalism developed for NSs remains (approximately) valid. For simplicity, we consider only $l=2$ perturbations for the remainder of this section. The generalization of these results to arbitrary $l$ is detailed in Ref.~\cite{Damour:2009vw}.

As shown in Eqs.~\eqref{eq:BackAsymptotics} and~\eqref{eq:PertAsymptotics}, very far from the center of the BS, the system approaches vacuum exponentially. Neglecting the vanishingly small contributions from the scalar field, the metric perturbation reduces to the general form
\begin{align}
h_0^{\rm vac}=c_1 \hat{Q}_{22}(x)+c_2 \hat{P}_{22}(x)+\Order\left[\left(\phi_0\right)^1,\left( \phi_1\right)^1\right],\label{eq:VacuumSolution}
\end{align}
where we have defined ${x\equiv r/M-1}$, $\hat{P}_{22}$ and $\hat{Q}_{22}$ are the associated Legendre functions of the first and second kind, respectively,  normalized as in Ref.~\cite{Damour:2009vw} such that $\hat{P}_{22}\sim x^{2}$ and $\hat{Q}_{22}\sim 1/x^{3}$ when $x\rightarrow \infty$. The coefficients $c_1$ and $c_2$ are the same as in Eq.~\eqref{eq:h0Asymptotics}.

In the BS's local asymptotic rest frame, the metric far from the star's center takes the form \cite{Thorne:1997kt}
\begin{align}
\begin{split}
\bar{g}_{00}=&-1+\frac{2 M}{r}+\frac{3 \mathcal{Q}_{ij}}{r^3}\left(n^i n^j-\frac{1}{3}\delta^{ij}\right)+\Order\left(\frac{1}{r^4}\right)\\
&-\mathcal{E}_{ij} x^i x^j +\Order\left(r^3 \right)+\Order\left[\left(\phi_0\right)^1,\left( \phi_1\right)^1\right],\label{eq:LocalAsymptoticFrame}
\end{split}
\end{align}
where ${n^i=x^i/r}$, $\mathcal{E}_{ij}$ is the external tidal field, and $\mathcal{Q}_{ij}$ is the induced quadrupole moment. Working to linear order in $\mathcal{E}_{ij}$, the tidal deformability $\lambda_\text{Tidal}$ is defined such that
\begin{align}
\mathcal{Q}_{ij}=-\lambda_\text{Tidal} \mathcal{E}_{ij}. \label{eq:lambdaTidalDef}
\end{align}
For our purposes, it will be convenient to instead work with the dimensionless quantity
\begin{align}
\Lambda\equiv \frac{\lambda_\text{Tidal}}{M^5}. \label{eq:LambdaDef}
\end{align} 
Comparing Eqs.~\eqref{eq:VacuumSolution} and~\eqref{eq:LocalAsymptoticFrame}, one finds that the tidal deformability can be extracted from the asymptotic behavior of $h_0$ using
\begin{align}
\Lambda=\frac{ c_1}{3 c_2}. \label{eq:Lambdaofci}
\end{align}
 
From Eq.~\eqref{eq:VacuumSolution}, the logarithmic derivative 
\begin{align}
y\equiv \frac{d\log h_0}{d\log r}= \frac{r h_0'}{h_0}, \label{eq:ydef}
\end{align}
takes the form
\begin{align}
y(x)=(1+x)\frac{3 \Lambda \hat{Q}_{2 2}'(x)+ \hat{P}_{2 2}'(x)}{3 \Lambda \hat{Q}_{2 2}(x)+ P_{2 2}(x)},\label{eq:yVacuum}
\end{align}
or equivalently
\begin{align}
\Lambda=-\frac{1}{3}\left(\frac{(1+x)\hat{P}_{2 2}'(x)- y(x)\hat{P}_{2 2}(x)}{(1+x) \hat{Q}_{2 2}'(x)-y(x)\hat{Q}_{2 2}(x)}\right).\label{eq:LambdaY}
\end{align}
Starting from a numerical solution to the perturbation equations~\eqref{eq:Perturbationh} and~\eqref{eq:Perturbationphi}, one obtains the deformability $\Lambda$ by first computing $y$ from Eq.~\eqref{eq:ydef} and then evaluating Eq.~\eqref{eq:LambdaY} at a particular extraction radius $x_\text{Extract}$ far from the center of the BS. Details concerning the numerical extraction are described in Sec.~\ref{sec:NumericalMethods} below.

\section{Solving the background and perturbation equations}\label{sec:NumericalMethods}
The background equations~\eqref{eq:TOVu}--\eqref{eq:TOVphi}
  and perturbation
  equations~\eqref{eq:Perturbationh}--\eqref{eq:Perturbationphi} form
  systems of coupled ordinary differential
  equations. These equations can be simplified by
rescaling the coordinates and fields by $\mu$ (the mass of the boson
field). To ease the comparison with previous work, we extend the
definitions given in Ref. \cite{Macedo:2013jja}: for massive BSs, we
use
\begin{align}
\begin{split}
r \rightarrow& \frac{m_\text{Planck}^2\tilde{r}}{\mu},\qquad m(r)\rightarrow \frac{m_\text{Planck}^2\tilde{m}(\tilde{r})}{\mu},\\
\lambda \rightarrow & \frac{8 \pi \mu^2 \tilde{\lambda}}{m_\text{Planck}^2},\qquad \omega \rightarrow \frac{\mu \tilde{\omega}}{m_\text{Planck}^2}, \\
\phi_0(r)\rightarrow& \frac{m_\text{Planck} \tilde{\phi}_0(\tilde{r})}{(8 \pi)^{1/2}}, \qquad  \phi_1(r)\rightarrow \frac{m_\text{Planck}^2\tilde{\phi}_1(\tilde{r})}{\mu (8 \pi)^{1/2}},\label{eq:Rescalings}
\end{split}
\end{align}
while for solitonic BSs, we use

\begin{align}
\begin{split}
r \rightarrow& \frac{m_\text{Planck}^2\tilde{r}}{\tilde\sigma_0 \mu},\qquad m(r)\rightarrow \frac{m_\text{Planck}^2\tilde{m}(\tilde{r})}{\tilde \sigma_0 \mu},\\
\sigma_0 \rightarrow & \frac{ m_\text{Planck} \tilde{\sigma}_0}{(8\pi)^{1/2}},\qquad \omega \rightarrow \frac{\tilde{\sigma}_0 \mu \tilde{\omega}}{m_\text{Planck}^2}, \\
\phi_0(r)\rightarrow& \frac{\sigma_0 \tilde{\phi}_0(\tilde{r})}{(2)^{1/2}}, \qquad  \phi_1(r)\rightarrow \frac{m_\text{Planck}^2\tilde{\phi}_1(\tilde{r})}{(16 \pi)^{1/2} \mu},\label{eq:RescalingsSolitonic}
\end{split}
\end{align}
where factors of the Planck mass have been restored for clarity.

Finding solutions with the proper asymptotic behavior
[Eqs.~\eqref{eq:BackAsymptotics} and~\eqref{eq:PertAsymptotics}]
requires one to specify boundary conditions at both $\tilde{r}=0$ and
$\tilde{r}=\infty$. To impose these boundary conditions precisely, we
integrate over a compactified radial coordinate
\begin{align}
\zeta=\frac{\tilde{r}}{N+\tilde{r}},
\end{align}
as is done in Ref.~\cite{Lai:2004fw}, where $N$ is a parameter tuned
so that exponential tails in the variables $\tilde{\phi}_0$ and $\tilde{\phi}_1$ [see Eqs.~\eqref{eq:BackAsymptotics} and~\eqref{eq:PertAsymptotics}]
begin near the center of the domain $\zeta\in[0,1]$. For massive BSs,
we use $N$ ranging from 20 to 60 depending on the body's compactness;
for solitonic BSs we use $N$ between 1 and 10.

Ground-state solutions to the background
equations~\eqref{eq:TOVu}--\eqref{eq:TOVphi} can be completely
parameterized by the central scalar field $\tilde\phi_0^{(c)}$ and frequency
$\tilde{\omega}$. To determine the ground state frequency, we formally
promote $\tilde{\omega}$ to an unknown constant function of
$\tilde{r}$ and simultaneously solve both the background equations and
\begin{align}
\tilde{\omega}'(\tilde{r})=0.
\end{align}
We impose the following boundary conditions on this combined system:
\begin{align}
\begin{split}
u(0)=&0, \qquad \tilde{\phi}_0(0)=\tilde{\phi}_0^{(c)}, \qquad \tilde{\phi}_0'(0)=0,\\
v(\infty)=&0, \qquad \tilde{\phi}_0(\infty)=0.
\end{split}
\end{align}
Here, the inner boundary conditions ensure regularity at the
origin, and the outer conditions guarantee asymptotic flatness.

The background and pertrubation equations are stiff, and therefore the shooting
  techniques usually used to solve two-point boundary value
  problems require signficant fine-tuning to converge to a
  solution~\cite{Macedo:2013jja}. To avoid these difficulties, we use a standard relaxation
  algorithm that more easily finds a solution given a
  reasonable initial guess~\cite{press1992numerical}. Once a solution is found for a particular choice of the central scalar field $\tilde\phi_0^{(c)}$ and scalar coupling (i.e., $\lambda$ for massive BSs or $\sigma_0$ for solitonic BSs), this solution can be used as an initial guess to obtain nearby solutions. By iterating this process, one can efficiently generate many BS configurations.

After finding a background solution, we solve the perturbation
equations~\eqref{eq:Perturbationh} and~\eqref{eq:Perturbationphi}. To
improve numerical behavior of the perturbation equations near the
boundaries, we factor out the dominant $\tilde{r}$ dependence and
instead solve for
\begin{align}
\bar{h}_0(\tilde{r})\equiv h_0 \tilde{r}^{-2},\\
\bar{\phi}_1(\tilde{r})\equiv \tilde{\phi}_1 \tilde{r}^{-3}.
\end{align}
We employ the boundary conditions 
\begin{align}
\begin{split}
\bar{h}_0(0)&=\bar{h}_0^{(c)}, \qquad \bar{h}_0'(0)=0,\\
\bar{\phi}_1'(0)&=0, \qquad \bar{\phi}_1(\infty)=0,
\end{split}
\end{align}
where the normalization $\bar{h}_0^{(c)}$ is an arbitrary non-zero constant.

\begin{figure}
  \includegraphics[width=\columnwidth]{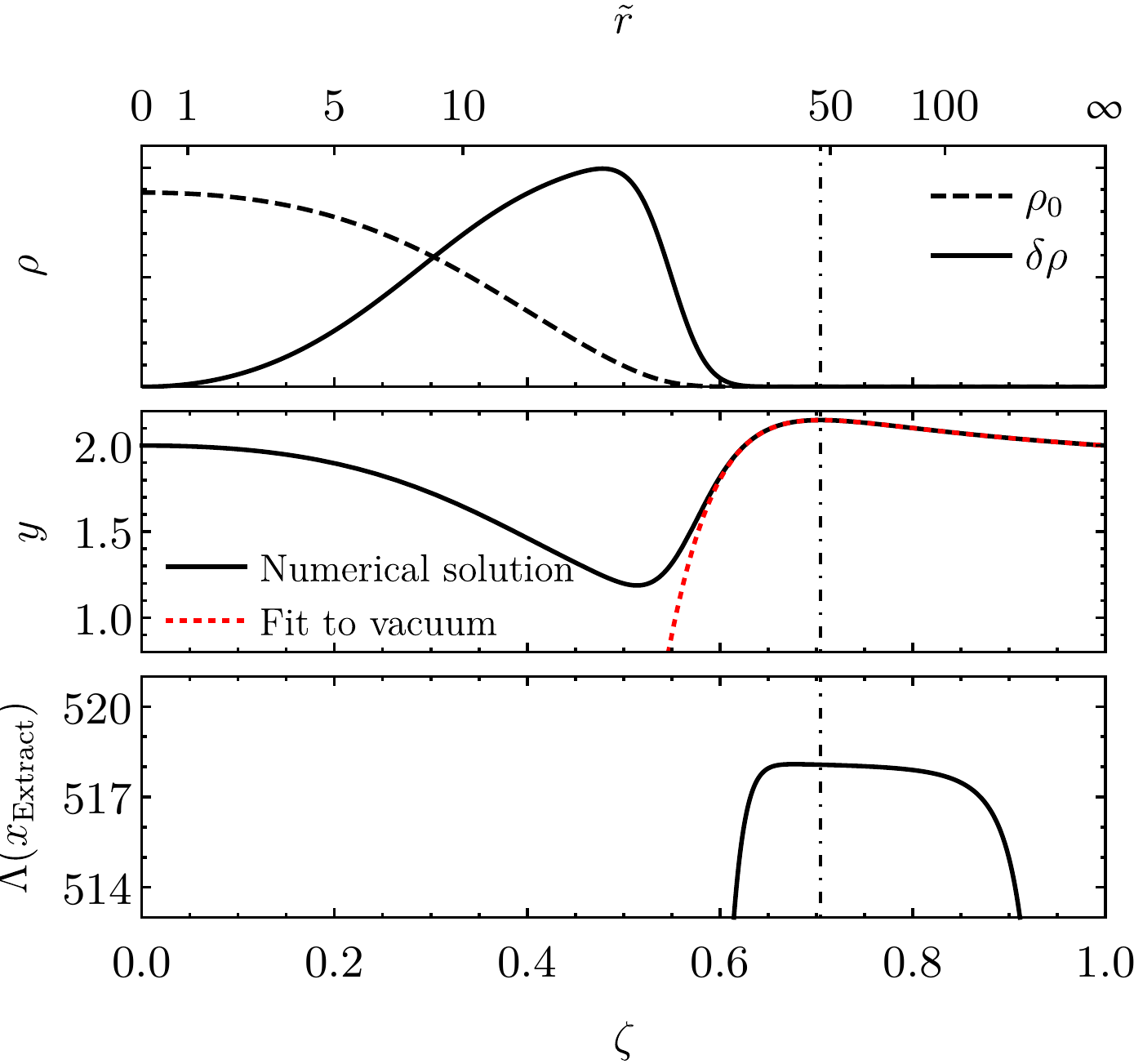}
\caption{Perturbations of a massive BS as a function of rescaled coordinate $\tilde{r}$ and compactified coordinate $\zeta$ for a star of mass ${M=3.78 m_\text{Planck}^2/\mu}$ with coupling ${\tilde{\lambda}=300}$. \textit{Top panel:} The background density $\rho_0$ (dashed) and its first-order perturbation $\delta \rho$ (solid), rescaled to fit on the same plot. \textit{Middle panel:} Logarithmic derivative $y$ of the metric perturbation. The tidal deformability $\Lambda$ is calculated using the numerically computed solution (black) at the peak of $y$ (dot-dashed vertical line). Using this value for $\Lambda$, we plot corresponding expected behavior in vacuum (red) as given by Eq.~\eqref{eq:yVacuum}. \textit{Bottom panel:} Tidal deformability computed from Eq.~\eqref{eq:LambdaY} as a function of extraction radius $x_\text{Extract}$.}\label{fig:RhoYZeta}
\end{figure}

Finally, we compute the tidal deformability using Eq.~\eqref{eq:LambdaY} in the nearly vacuum region $x\gg1$. At very large distances, the exponential falloff of $\phi_0$ and $\phi_1$ is difficult to resolve numerically. This numerical error propagates through the computation of the tidal deformability in  Eq.~\eqref{eq:LambdaY} for very large values of $x$. We find that extracting $\Lambda$ at smaller radii provides more numerically stable results, with a typical variation of ${\sim 0.1\%}$ for different choices of extraction radius $x_\text{Extract}$. For consistency, we extract $\Lambda$ at the radius at which $y$ attains its maximum.

Figure~\ref{fig:RhoYZeta} demonstrates our procedure for computing the tidal deformability. The background and perturbation equations are solved for a massive BS with a coupling of $\tilde{\lambda}=300$ using a compactified coordinate with $N=20$. The profile of the effective density $\rho$, decomposed into its background value $\rho_0$ and first order correction $\delta \rho$, is shown in the top panel for a star of mass $3.78m_\text{Planck}^2/\mu$. Note that the magnitude of the perturbation is proportional to the strength of the external tidal field; to improve readability, we have scaled $\delta \rho$ to match the size of $\rho_0$. 

The middle panel of Fig.~\ref{fig:RhoYZeta} shows the computed logarithmic derivative $y$ across the entire spacetime (black). We calculate the deformability with Eq.~\eqref{eq:LambdaY} using the peak value of $y$, located at the dot-dashed line. Comparing with the top panel, one sees that the scalar field is negligible in this region, justifying our use of formulae valid in vacuum. The bottom panel depicts the typical variation of $\Lambda$ computed at different locations $x_\text{Extract}$---our procedure yields consistent results provided one works reasonably close to the edge of the BS. As a check, we insert the computed value of $\Lambda$ back into the vacuum solution for $y$ given in Eq.~\eqref{eq:yVacuum}, plotted in red in the middle panel. As expected, this curve closely matches the numerically computed solution at large radii, but deviates upon entering a region with non-negligible scalar field.

\begin{figure*}
  \includegraphics[width=\textwidth]{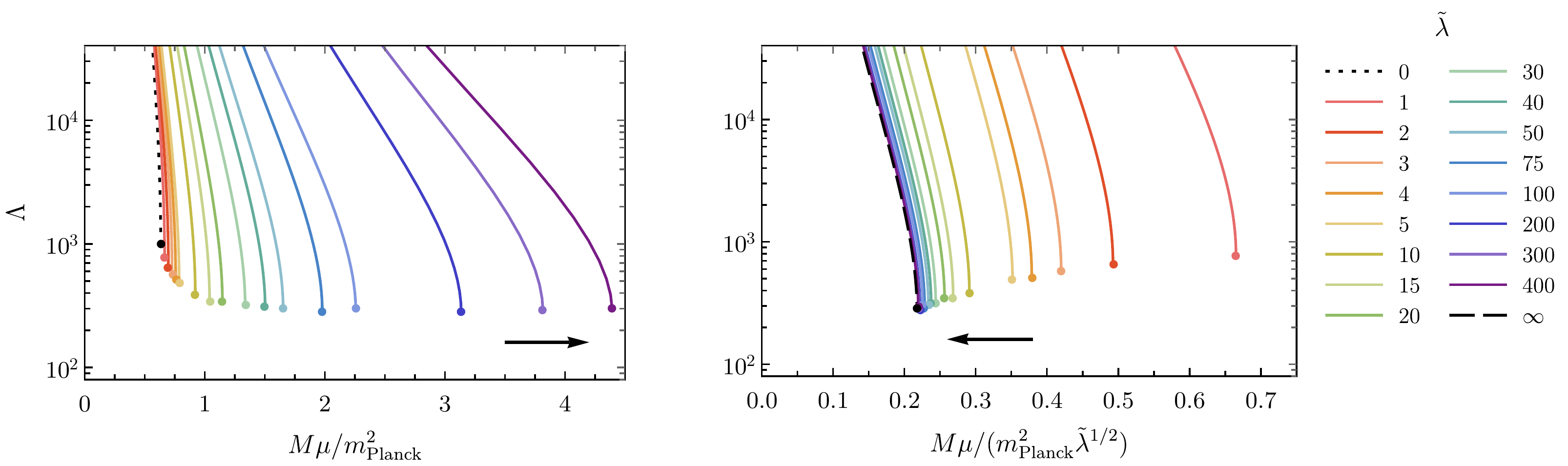}
\caption{Dimensionless tidal deformability of a massive BS as a function of mass in units of (left) $m_\text{Planck}^2/\mu$ and (right) $m_\text{Planck}^2 \tilde{\lambda}^{1/2}/\mu$. For each value of $\tilde{\lambda}$, the most compact stable configuration is highlighted with a colored dot. The arrows indicate the direction towards the strong-coupling regime, i.e. of increasing $\tilde\lambda$.}\label{fig:LambdaMtil}
\end{figure*}

\section{Results}\label{sec:Results}

\subsection{Massive Boson Stars}\label{sec:MassiveResults}
The dimensionless tidal deformability of massive BSs is given as a function of the rescaled total mass $\tilde{M}$ [defined as in Eq.~\eqref{eq:Rescalings}] in the left panel of Fig.~\ref{fig:LambdaMtil}. The deformability in the weak-coupling limit $\tilde{\lambda}=0$ is given by the dotted black curve; this limit corresponds to the mini BS model considered in Ref.~\cite{Mendes:2016vdr}.\footnote{In Ref.~\cite{Mendes:2016vdr}, the authors computed the quantity $k_\text{BS}$, related to the quantity $\Lambda$ presented here by ${k_\text{BS}=\Lambda M^{10}}$. The quantity $k_2^E$, computed in Ref.~\cite{Cardoso:2017cfl} for mini, massive, and solitonic BSs, is related to $\Lambda$ by ${k_2^E=(4 \pi/5)^{1/2}\Lambda}$.} One finds that the tidal deformability of the most massive stable star (colored dots) decreases from $\Lambda\sim900$ in the weak-coupling limit towards $\Lambda\sim 280$ as $\tilde{\lambda}$ is increased. For large values of $\tilde\lambda$, the deformability exhibits a universal relation when written in terms of the rescaled mass $\tilde M / \tilde \lambda^{1/2}$ in the sense that the results for large $\tilde \lambda$ rapidly approach a fixed curve as the coupling strength increases. This convergence towards the $\tilde\lambda=\infty$ relation is illustrated in the right panel of Fig.~\ref{fig:LambdaMtil}, in which the x-axis is rescaled by an additional factor of $\tilde{\lambda}^{1/2}$ relative to the left panel; in both panels, we have added black arrows to indicate the direction of increasing $\tilde\lambda$. Employing this rescaling of the mass, we compute the relation $\Lambda(\tilde{M},\tilde{\lambda})$ in the strong-coupling limit $\tilde{\lambda}\rightarrow \infty$ below. The tidal deformability in this limit is plotted in Fig.~\ref{fig:LambdaMtil} with a dashed black curve. 

The gap in tidal deformability between BSs, for which the lowest values are $\Lambda\gtrsim 280$, and NSs, where for soft equations of state and large masses $\Lambda \gtrsim 10$, can be understood by comparing the relative size or compactness $C=M/R$ of each object. From the definitions~\eqref{eq:lambdaTidalDef} and~\eqref{eq:LambdaDef}, one expects the tidal deformability to scale as $\Lambda\propto 1/C^5$. In the strong-coupling limit, stable massive BSs can attain a compactness of ${C_\text{max} \approx 0.158}$; note that in the exact strong-coupling limit $\tilde\lambda=\infty$, BSs develop a surface, and thus their compactness can be defined unambiguously.  A NS of comparable compactness has a tidal deformability that is only~${\sim 0 \enDash 25\%}$ larger than that of BSs. However, NS models predict stable stars with approximately twice the compactness that can be attained by massive BSs, and thus, their minimum tidal deformability is correspondingly much lower.

As argued in Sec.~\ref{sec:BSbasics}, the strong-coupling limit of
massive BSs is the most plausible model investigated in this
paper from an effective field theory perspective. We analyze the tidal deformability in this
  limit in greater detail.  To study the strong-coupling limit of
$\tilde{\lambda}\rightarrow \infty$, we employ a different set of
rescalings introduced, first in Ref. \cite{Colpi:1986ye}:
\begin{align}
\begin{split}
r \rightarrow& \frac{m_\text{Planck}^2\tilde{\lambda}^{1/2}\hat{r}}{\mu},\qquad m(r)\rightarrow \frac{m_\text{Planck}^2\tilde{\lambda}^{1/2}\hat{m}(\hat{r})}{\mu},\\
\lambda \rightarrow & \frac{8 \pi \mu^2 \tilde{\lambda}}{m_\text{Planck}^2},\qquad \omega \rightarrow \frac{\mu \hat{\omega}}{m_\text{Planck}^2},\\
 \phi_0(r)\rightarrow& \frac{m_\text{Planck}\hat{\phi}_0(\hat{r})}{(8 \pi \tilde{\lambda})^{1/2}} , \qquad  \phi_1(r)\rightarrow \frac{m_\text{Planck}^2\hat{\phi}_1(\hat{r})}{\mu(8 \pi)^{1/2}},
\end{split}
\end{align}
where we have kept the previous notation for $\tilde{\lambda}$ to emphasize that it is the same quantity as defined in Eq.~\eqref{eq:Rescalings}.

Keeping terms only at leading order in $\tilde\lambda^{-1}\ll 1$, Eqs.~\eqref{eq:TOVu}--\eqref{eq:TOVphi} become
\begin{gather}
e^{-u}\left(-\frac{u'}{\hat{r}}+\frac{1}{\hat{r}^2}\right)-\frac{1}{\hat{r}^2}=-2\hat\phi_0^2-\frac{3\hat\phi_0^4}{2},\label{eq:TOVuLargeLambda}\\
e^{-u}\left(\frac{v'}{\hat{r}}+\frac{1}{\hat{r}^2}\right)-\frac{1}{\hat{r}^2}=\frac{\hat\phi_0^4}{2},\label{eq:TOVvLargeLambda}\\
\hat\phi_0=\left(\hat\omega^2 e^{-v}-1\right)^{1/2},
\end{gather}
where a prime denotes differentiation with respect to $\hat{r}$. Note that in particular, Eq.~\eqref{eq:TOVphi} becomes an algebraic equation, reducing the system to a pair of first order differential equations.

Turning now to the perturbation equations, we use these rescalings and find that to leading order in $\tilde\lambda^{-1}$, Eqs.~\eqref{eq:Perturbationh} and \eqref{eq:Perturbationphi} become
\begin{widetext}
\begin{align}
\begin{split}
&h_0''+\frac{e^{u}h_0'}{\hat r}\left[\frac{\hat{r}^2}{2} \left(1-e^{-2 v}\hat{\omega}^4\right)+e^{-u}+1 \right] \\
&-\frac{e^u h_0}{\hat{\hat{r}}^2}\left[\frac{\hat{r}^4 e^{u}}{4}  \left(1-e^{-v}\hat{\omega} ^2\right)^4
+\hat{r}^2 \left(e^{u}(1-e^{-v}\hat{\omega}^2)^2+10 e^{-v} \hat{\omega}^2(1-e^{-v} \hat{\omega}^2)-2\right)+e^{u}(1-e^{-u})^2+l(l+1)\right]=0 ,
\end{split}\label{eq:PerturbationhLargeLambda} \\
&\hat\phi_1=\frac{h_0 \hat r \left(1+\hat\phi_0^2\right)}{2 \hat\phi_0}.\label{eq:PerturbationphiLargeLambda}
\end{align}
\end{widetext}
As with the background fields, the equation for the scalar field $\hat{\phi}_1$ becomes algebraic in this limit. Note that the scalar perturbation diverges as one approaches the surface of the BS, defined as the shell on which $\hat{\phi}_0$ vanishes. Nevertheless, the metric perturbation $h_0$ remains smooth over this surface.

We integrate the simplified background equations~\eqref{eq:TOVuLargeLambda} and~\eqref{eq:TOVvLargeLambda} and then the perturbation equation~\eqref{eq:PerturbationhLargeLambda} using Runge-Kutta methods. We compute the tidal deformability using Eq.~\eqref{eq:LambdaY} evaluated at the surface of the BS, and plot the results in the right panel of Fig.~\ref{fig:LambdaMtil} (dashed black).

\subsection{Solitonic boson stars}

\begin{figure*}
  \includegraphics[width=\textwidth]{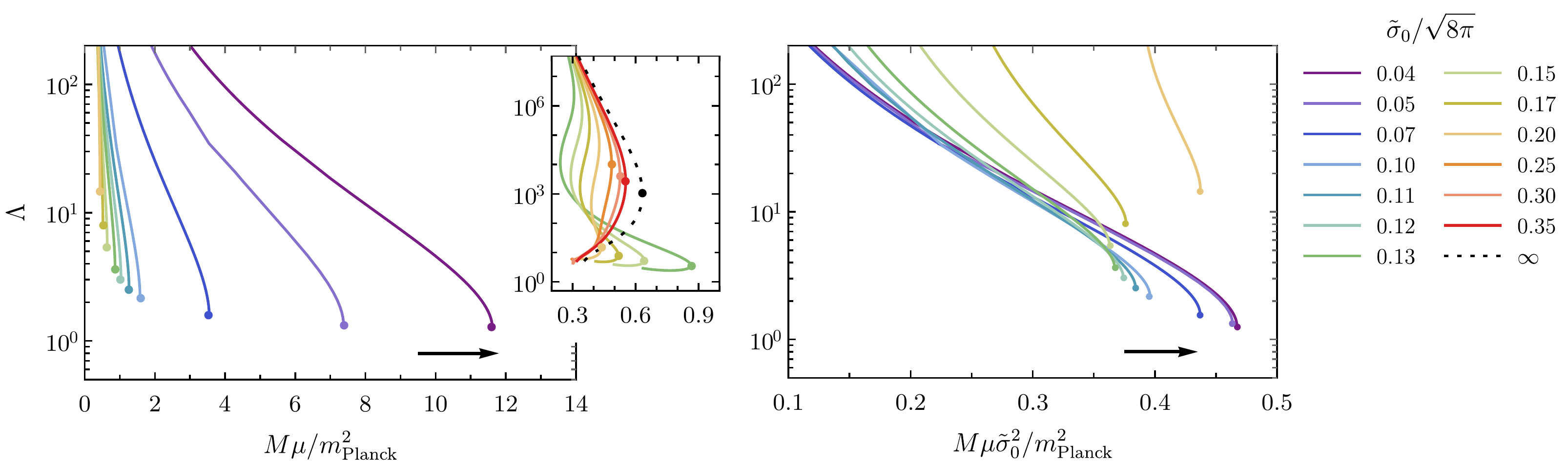}
\caption{Dimensionless tidal deformability as a function BS mass in units of (left) $m_\text{Planck}^2/\mu$ and (right) $m_\text{Planck}^2 /(\mu \tilde{\sigma}_0^2)$. For each value of $\tilde{\sigma}_0$, the most compact stable BS is highlighted with a colored dot. The inset plot in the left panel shows both stable and unstable configurations over a larger range of $\Lambda$ to illustrate the weak-coupling limit $\tilde\sigma_0\rightarrow \infty$ (dotted black). The arrows indicate the direction towards the strong-coupling regime, i.e. of decreasing $\tilde\sigma_0$; while not plotted explicitly, the strong-coupling limit $\tilde\sigma_0\rightarrow 0$ corresponds the accumulation of curves in the right panel in the direction of the arrow.}\label{fig:LambdaMtilSolitonic}
\end{figure*}

The dimensionless tidal deformability of solitonic BSs is given as a function of the mass in Fig.~\ref{fig:LambdaMtilSolitonic}. As in Fig.~\ref{fig:LambdaMtil}, the colored dots highlight the most massive stable configuration for different choices of the scalar coupling $\tilde\sigma_0$. To aid comparison with the massive BS model, in the left panel we rescale the mass by an additional factor of $\tilde{\sigma}_0$ relative to the definition of $\tilde{M}$ in Eq.~\eqref{eq:RescalingsSolitonic}.

When the coupling $\tilde\sigma_0$ is strong, solitonic BSs can manifest two stable phases that can be smoothly connected through a sequence of unstable configurations \cite{Kleihaus:2011sx}. The large plot in the left panel only shows stable configurations on the more compact branch of configurations. In the weak-coupling limit $\tilde{\sigma}_0\rightarrow\infty$, solitonic BSs reduce to the free field model considered in Ref.~\cite{Mendes:2016vdr}. To illustrate this limit, we show in the smaller inset the tidal deformability for both phases of BSs as well as the unstable configurations that bridge the two branches of solutions. The weak-coupling limit is depicted with a dotted black curve. We find that the tidal deformability of the less compact phase of BSs smoothly transitions from $\Lambda\rightarrow \infty$ in the strong-coupling limit ($\tilde\sigma_0\rightarrow 0$)\footnote{In the exact strong-coupling limit $\tilde\sigma_0 = 0$, this diffuse phase of solitonic BSs vanishes \cite{Friedberg:1986tq}. However, the tidal deformability of this branch of BS configurations can be made arbitrarily large by choosing $\tilde\sigma_0$ to be sufficiently small.} to $\Lambda\sim900$ in the weak-coupling limit ($\tilde\sigma_0\rightarrow \infty$). Because their tidal deformability is so large, diffuse solitonic BSs of this kind would not serve as effective BH mimickers, and we will not discuss them for the remainder of this paper. However, it should be noted that only this phase of stable configurations exists when $\sigma_0\gtrsim 0.23\,m_\text{Planck}$.

Focusing now on the more compact phase of solitonic  BSs, one finds that the tidal deformability of the most massive stable star (colored dots) decreases towards $\Lambda\sim 1.3$ as $\tilde{\sigma}_0$ is decreased. As before, the relation between a rescaled mass and $\Lambda$ approaches a finite limit in the strong coupling limit. We illustrate this in the right panel of Fig.~\ref{fig:LambdaMtilSolitonic} by rescaling the mass by an additional factor of $\tilde{\sigma}_0^{-1}$ relative to the definition in Eq.~\eqref{eq:RescalingsSolitonic}. While we do not examine the exact strong-coupling limit $\tilde\sigma_0\rightarrow 0$ here, we find that the minimum deformability has converged to within a few percent of $\Lambda=1.3$ for ${0.03\,m_\text{Planck}\leq\sigma_0\leq0.05\, m_\text{Planck}}$.

\subsection{Fits for the relation between $M$ and $\Lambda$}\label{sec:Fits}

In this section we provide fits to our results for practical use in data analysis studies, focusing on the regime that is the most relevant region of the parameter space for BH and NS mimickers.

For massive BS, it is convenient to express the fit in terms of the variable 
\begin{equation}
w = \frac{1}{1 + \tilde \lambda / 8} ,
\end{equation}
which provides an estimate of the maximum mass in the weak-coupling limit $ \tilde{M}_\text{max} \approx 2 /( \pi \sqrt{w})$ \cite{Mielke:2000mh}
and has a compact range $0 \leq w \leq 1$. A fit for massive BSs that is accurate\footnote{The accuracy quoted here corresponds to the prediction for the mass at fixed $\Lambda$ and coupling constant. The error in $\Lambda$ at a fixed mass can be much larger, because $\Lambda$ has a large gradient when varying the mass, which even diverges at the maximum mass. The applicability of our fits must be judged by the accuracy with which the masses can be measured from a GW signal.} to
$\sim 1\%$ for $\Lambda \leq 10^5$ and up to the maximum mass is given by
\begin{equation}
\begin{split}
&\sqrt{w} \tilde M = \left[ -0.529 + \frac{22.39}{\log \Lambda} - \frac{143.5}{(\log \Lambda)^2} + \frac{305.6}{(\log \Lambda)^3} \right] w \\
&+ \left[ -0.828 + \frac{20.99}{\log \Lambda} - \frac{99.1}{(\log \Lambda)^2} + \frac{149.7}{(\log \Lambda)^3} \right] (1-w) .
\end{split}
\end{equation}
The maximum mass where
the BSs become unstable can be obtained from the extremum of this fit, which also determines
the lower bound for $\Lambda$.

In the solitonic case, a global fit for the tidal deformability for all possible values of $\sigma_0$ is difficult to obtain due
to qualitative differences between the weak- and strong-coupling regimes.
However, small values of $\sigma_0$ are most interesting, since they allow for the widest range for
the tidal deformability and compactness.
A fit for $\sigma_0 = 0.05\,m_\text{Planck}$ accurate to better than $1\%$ and valid for
$\Lambda \leq 10^4$ (and again up to the maximum mass) reads
\begin{equation}
\log (\sigma_0 \tilde M) = -30.834 + \frac{1079.8}{\log\Lambda + 19} - \frac{10240}{(\log\Lambda + 19)^2} .
\end{equation}
This fit is expected to be accurate for ${0 \leq \sigma_0 \lesssim 0.05\, m_\text{Planck}}$, i.e., including the strong coupling limit $\sigma_0 = 0$, within a few percent. Notice that this fit remains valid through tidal deformabilities of the same magnitude as that of NSs.

\section{Prospective constraints} \label{sec:Constraints}
\subsection{Estimating the precision of tidal deformability measurements}
Gravitational-wave detectors will be able to probe the structure of compact objects through their tidal interactions in binary systems, in addition to effects seen in the merger and ringdown phases. In this section, we discuss the possibility of distinguishing BSs from NSs and BHs using only tidal effects. We emphasize that our results in this section are based on several approximations and should be viewed only as estimates that provide lower bounds on the errors and can be used to identify promising scenarios for future studies with Bayesian data analysis and improved waveform models.

The parameter estimation method based on the Fisher information matrix is discussed in detail in Ref.~\cite{Cutler:1994ys}. This approximation yields only a lower bound on the errors that would be obtained from a Bayesian analysis. We assume that a detection criterion for a GW signal $h(t;\boldsymbol{\theta})$ has been met, where $\boldsymbol{\theta}$ are the parameters characterizing the signal: the distance $D$ to the source, time of merger $t_c$, five positional angles on the sky, plane of the orbit, orbital phase at some given time $\phi_c$, as well as a set of intrinsic parameters such as orbital eccentricity, masses, spins, and tidal parameters of the bodies. Given the detector output $s=h(t)+n$, where $n$ is the noise, the probability $p(\boldsymbol{\theta}|s)$ that the signal is characterized by the parameters $\boldsymbol{\theta}$ is
\begin{equation}
p(\boldsymbol{\theta}|s)\propto p^{(0)} e^{-\frac{1}{2} (h(\boldsymbol{\theta})-s|h(\boldsymbol{\theta})-s)}, \label{eq:prob}
\end{equation}
where $ p^{(0)} $ represents a priori knowledge. Here, the inner product $(\cdot|\cdot)$ is determined by the statistical properties of the noise and is given by 
\begin{equation}
(h_1|h_2)=2\int_0^\infty \frac{\tilde{h}_1^*(f)\tilde{h}_2(f)+\tilde{h}_2^*(f)\tilde{h}_1(f)}{S_n(f)}df, \label{eq:innerprod}
\end{equation}
where $S_n(f)$ is the spectral density describing the Gaussian part of the detector noise. For a measurement, one determines the set of best-fit parameters $\hat{\boldsymbol{\theta}}$ that maximize the probability distribution function~\eqref{eq:prob}. In the regime of large signal-to-noise ratio $\text{SNR}=\sqrt{(h|h)}$, for a given incident GW in different realizations of the noise, the probability distribution $p(\boldsymbol{\theta}|s)$ is approximately given by
\begin{equation}
p(\boldsymbol{\theta}|s)\propto p^{(0)} e^{-\frac{1}{2} \Gamma_{ij}\Delta \theta^i \Delta \theta^j}, \label{eq:pGamma}
\end{equation}
where 
\begin{equation}
\Gamma_{ij}=\left(\frac{\partial h}{\partial \theta^i}\biggl \lvert \frac{\partial h}{\partial \theta^j} \right) , \label{eq:Gamma}
\end{equation}
is the so-called Fisher information matrix. For a uniform prior $p^{(0)}$, the distribution~\eqref{eq:pGamma} is a multivariate Gaussian with covariance matrix 
$\Sigma^{ij}=(\boldsymbol{\Gamma}^{-1})^{ij}$
and the root-mean-square measurement errors in $\theta^i$ are given by 
\begin{equation}
\sqrt{\langle (\Delta \theta^i)^2\rangle}=\sqrt{(\boldsymbol{\Gamma}^{-1})^{ii}},
\end{equation}
where angular brackets denote an average over the probability distribution function~\eqref{eq:pGamma}.

We next discuss the model $\tilde h(f, \boldsymbol{\theta})$ for the signal. For a binary inspiral, the Fourier transform of the dominant mode of the signal has the form 
\begin{equation}
\tilde h(f, \boldsymbol{\theta})={\cal A}(f,\boldsymbol{\theta}) e^{i\psi(f,\boldsymbol{\theta})}.
\end{equation}
Using a PN expansion and the stationary-phase approximation (SPA), the phase $\psi$ is computed from the energy balance argument by solving 
\begin{equation}
\frac{d^2\psi}{d\Omega^2}=\frac{2}{d\Omega/dt}=2 \frac{(dE/d\Omega)}{\dot E_{\rm GW}}, \ \ \ \ \label{eq:ddpsidomega}
\end{equation}
where $E$ is the energy of the binary system, $\dot E_{\rm GW}$ is the energy flux in GWs, and $\Omega=\pi f$ is the orbital frequency. The result is of the form
\begin{align}
\psi=&\frac{3}{128(\pi{\cal M}f)^{5/2}}\bigg[1+\alpha_{\rm 1PN}(\nu) x+\ldots \qquad \; \; \; \; \; \nonumber\\
&\qquad \; \; \; \; \; +\left(\alpha_{\rm tidal}^{\rm Newt}+\alpha_{\rm 5PN}(\nu)\right) x^5+{\cal O}(x^6)\bigg], \label{eq:psiform}
\end{align}
with $x=(\pi M f)^{2/3}$, $M=m_1+m_2$, $\nu=m_1m_2/M^2$, ${\cal M}=\nu^{3/5}M$, and the dominant tidal contribution is
\begin{equation}
\alpha_{\rm tidal}^{\rm Newt}=-\frac{39}{2}\tilde \Lambda.
\end{equation}
Here, $\tilde \Lambda$ is the weighted average of the individual tidal deformabilities, given by
\begin{equation}
\tilde \Lambda(m_1,m_2,\Lambda_1,\Lambda_2)=\frac{16}{13} \left[\left(1+12\frac{m_2}{m_1}\right)\frac{m_1^5}{M^5}\Lambda_1+\left(1\leftrightarrow 2\right)\right]. \label{eq:Lambdatildesimple}
\end{equation}
The phasing in Eq.~\eqref{eq:psiform} is known as the ``TaylorF2 approximant.''  Specifically, we use here the $3.5$PN point-particle terms~\cite{Blanchet:2013haa} and the 1PN tidal terms~\cite{Vines:2011ud}. At 1PN order, a second combination of tidal deformability parameters enters into the phasing in addition to $\tilde \Lambda$. This additional parameter vanishes for equal-mass binaries and will be difficult to measure with Advanced LIGO \cite{Favata:2013rwa,Wade:2014vqa}. For simplicity, we omit this term from our analysis.

The tidal correction terms in Eq.~\eqref{eq:psiform} enter with a high power of the frequency, indicating that most of the information on these effects comes from the late inspiral. This is also the regime where the PN approximation for the point-mass dynamics becomes inaccurate. To estimate the size of the systematic errors introduced by using the TaylorF2 waveform model in our analysis, we compare the model against predictions from a tidal EOB (TEOB) model. The accuracy of the TEOB waveform model has been verified for comparable-mass binaries through comparison with NR simulations; see, for example, Ref.~\cite{Hinderer:2016eia}. For our comparison, we use the same TEOB model as in Ref.~\cite{Hinderer:2016eia}. The point-mass part of this model---known as ``SEOBNRv2''---has been calibrated with binary black hole (BBH) results from NR simulations. The added tidal effects are adiabatic quadrupolar tides including tidal terms at relative 2PN order in the EOB Hamiltonian and 1PN order in the fluxes and waveform amplitudes. The SPA phase for the TEOB model is computed by solving the EOB evolution equations to obtain $\Omega(t)$, numerically inverting this result for $t(\Omega)$, and solving Eq.~\eqref{eq:ddpsidomega} to arrive at $\psi(\Omega)$.

\begin{figure}
\includegraphics[width=\columnwidth]{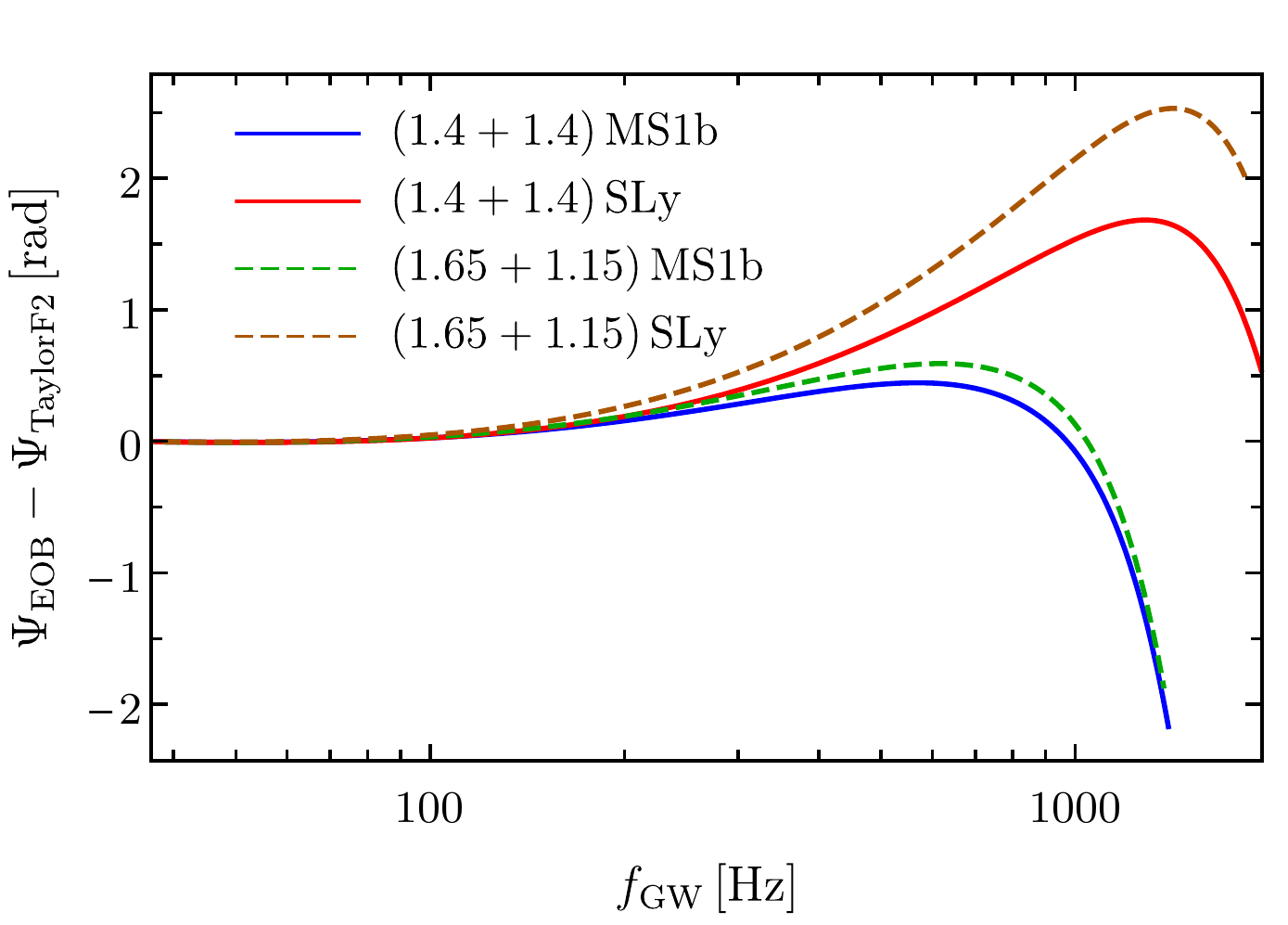}
\caption{Dephasing between the TEOB and tidal TaylorF2 models for non-spinning BNS systems including adiabatic quadrupolar tidal effects. The curves end at the prediction for the merger from NR simulations described in Ref.~\cite{Bernuzzi:2015rla}. The labels denote the masses (in units of $M_\odot$) and EoS of the NSs. }
\label{fig:EOBTaylorF2}
\end{figure}

Figure~\ref{fig:EOBTaylorF2} shows the difference in predicted phase from the TEOB model and the TaylorF2 model~\eqref{eq:psiform} for two nearly equal mass binary NS (BNS) systems. For our analysis, we consider two representative equations of state (EoS) for NSs: the relatively soft SLy model \cite{Douchin:2001sv} and the stiff MS1b EoS \cite{Mueller:1996pm}. Figure~\ref{fig:EOBTaylorF2} illustrates that the dephasing between the TaylorF2 and TEOB waveforms remains small compared to the size of tidal effects, which is on the order of $\gtrsim 20$ rad for MS1b $(1.4+1.4)M_\odot$. Thus, we conclude that the TaylorF2 approximant is sufficiently accurate for our purposes and leave an investigation of the measurability of tidal parameters with more sophisticated waveform models for future work.

Besides the waveform model, the computation of the Fisher matrix also requires a model of the detector noise. We consider here the Advanced LIGO Zero-Detuned High Power configuration \cite{Shoemaker:2010}. To assess the prospects for measurements with third-generation detectors we also use the ET-D \cite{Hild:2010id} and Cosmic Explorer \cite{Evans:2016mbw} noise curves.

To compute the measurement errors we specialize to the restricted set of signal parameters ${\boldsymbol{\theta}=\{\phi_c,t_c,{\cal M}, \nu, \tilde \Lambda\}}$. The extrinsic parameters of the signal such as orientation on the sky enter only into the waveform's amplitude and can be treated separately; they are irrelevant for our purposes. Spin parameters are omitted because the TaylorF2 approximant inadequately captures these effects and one would instead need to use a more sophisticated model such as SEOBNR. We restrict our analysis to systems with low masses $M\lesssim 12M_\odot$~\cite{Buonanno:2009zt} for which the merger occurs at frequencies $f_{\rm merger}>900$Hz so that the information is dominated by the inspiral signal. The termination conditions for the inspiral signal employed in our analysis are the predicted merger frequencies from NR simulations: for BNSs the formula from Ref.~\cite{Bernuzzi:2015rla}, and for BBH that from Ref.~\cite{Taracchini:2012ig}. 

From the tidal parameter $\tilde \Lambda$, we can obtain bounds on the individual tidal deformabilities. We adopt the convention that $m_1\geq m_2$. For any realistic, stable self-gravitating body, we expect an increase in mass to also increase the body's compactness. Because the tidal deformability scales as $\Lambda\propto 1/C^5$, we assume that $\Lambda_1\leq \Lambda_2$. At fixed values of $m_1,m_2,$ and $\tilde \Lambda$, the deformability of the more massive object $\Lambda_1$ takes its maximal value when it is exactly equal to $\Lambda_2$, i.e. when ${\tilde \Lambda=\tilde \Lambda(m_1,m_2,\Lambda_1,\Lambda_1)}$. Conversely, $\Lambda_2$ takes its maximal value when $\Lambda_1$ vanishes exactly so that ${\tilde \Lambda=\tilde \Lambda(m_1,m_2, 0, \Lambda_2)}$. Substituting the expression for $\tilde \Lambda$ from Eq.~\eqref{eq:Lambdatildesimple} and using that ${m_{1,2}=M(1\pm \sqrt{1-4\nu})/2}$ leads to the following bounds on the individual deformabilities
\begin{equation}
\label{eq:DeformabilityBounds}
\Lambda_1\leq g_1(\nu)\tilde{\Lambda}, \qquad
\Lambda_2\leq g_2(\nu)\tilde{\Lambda},
\end{equation}
where the functions $g_i$ are given by
\begin{subequations}
\begin{align}
g_1(\nu)\equiv&\frac{13}{16(1+7\nu-31 \nu^2)},\\
g_2(\nu)\equiv&\frac{13}{8\left[1+7\nu-31 \nu^2-\sqrt{1-4\nu}\left(1+9\nu-11\nu^2\right)\right]}. 
\end{align}
\end{subequations}
Thus, the expected measurement precision of $\nu$ and $\tilde\Lambda$ provide an estimate of the precision with which $\Lambda_1$ and $\Lambda_2$ can be measured through
\begin{subequations}\label{eq:DeltaLambdaDef}
\begin{align}
\Delta \Lambda_1&\leq\left[\left(g_1(\nu) \Delta \tilde\Lambda  \right)^2+\left(g_1'(\nu) \tilde\Lambda \Delta \nu \right)^2\right]^{1/2},\label{eq:DeltaLambda1Def}\\
\Delta \Lambda_2&\leq\left[\left(g_2(\nu) \Delta \tilde\Lambda  \right)^2+\left(g_2'(\nu) \tilde\Lambda \Delta \nu \right)^2\right]^{1/2},\label{eq:DeltaLambda2Def}
\end{align}
\end{subequations}
For simplicity, we have assumed in Eq.~\eqref{eq:DeltaLambdaDef} that the statistical uncertainty in $\nu$ and $\tilde{\Lambda}$ is uncorrelated. Note that for BBH signals, this assumption is unnecessary because $\tilde{\Lambda}=0$, and thus the second terms in Eqs.~\eqref{eq:DeltaLambda1Def} and~\eqref{eq:DeltaLambda2Def} vanish.

In the following subsections, we outline two tests to distinguish conventional GW sources from BSs and discuss the prospects of successfully differentiating the two with current- and third-generation detectors. First, we investigate whether one could accurately identify each body in a binary as a BH/NS rather than a BS. This test is only applicable to objects whose tidal deformability is significantly smaller than that of a BS, e.g., BHs and very massive NSs. For bodies whose tidal deformabilities are comparable to that of BSs, we introduce a novel analysis designed to test the slightly weaker hypothesis: can the binary system of BHs or NSs be distinguished from a binary BS (BBS) system? For both tests, we will assume that the true waveforms we observe are produced by BBH or BNS systems and then assess whether the resulting measurements are also consistent with the objects being BSs. In our analyses we consider only a single detector and assume that the sources are optimally oriented; to translate our results to a sky- and inclination-averaged ensemble of signals, one should divide the expected SNR by a factor of $\sqrt{2}$ and thus multiply the errors on $\Delta \tilde \Lambda$ by the same factor.

We consider two fiducial sets of binary systems in our analysis. First, we consider BBHs at a distance of 400 Mpc (similar to the distances at which GW150914 and GW151226 were observed~\cite{ Abbott:2016blz,TheLIGOScientific:2016pea}) with total masses in the range ${8 M_\odot \leq M \leq 12M_\odot}$. This range is determined by the assumption that the lowest BH mass is $4M_\odot$ and the requirement that the merger occurs at frequencies above $\sim 900$Hz so the information in the signal is dominated by the inspiral. The SNRs for these systems range from approximately 20 to 49  given the sensitivity of Advanced LIGO.  The second set of systems that we consider are BNSs at a distance of 200Mpc and with total masses ${2M_\odot \leq M\leq M_{\rm max}}$, where $M_{\rm max}$ is twice the maximum NS mass for each equation of state. The lower limit on this mass range comes from astrophysical considerations on NS formation~\cite{Belczynski:2011bn}. The BNS distance was chosen to describe approximately one out of every ten events within the expected BNS range of $\sim 300$ Mpc for Advanced LIGO and translates to $\text{SNR}\sim 12-22$ for the SLy equation of state.

\subsection{Distinguishability with a single deformability measurement}\label{sec:SingleMeasurement}

A key finding from Sec.~\ref{sec:Results} is that the tidal deformability is bounded below by $\Lambda \gtrsim 280$ for massive BSs and  $\Lambda \gtrsim 1.3$ for solitonic BSs. By comparison, the deformability of BHs vanishes exactly, i.e. $\Lambda=0$, whereas for nearly-maximal mass NSs, the deformability can be of order $\Lambda\approx \Order(10)$. Thus, a BH or high-mass NS could be distinguished from a massive BS provided that a measurement error of ${\Delta \Lambda\approx 200}$ can be reached with GW detectors. Similarly, to distinguish a BH from a solitonic BS requires a measurement precision of ${\Delta \Lambda\approx 1}$.

\begin{figure*}
\includegraphics[width=0.9\textwidth]{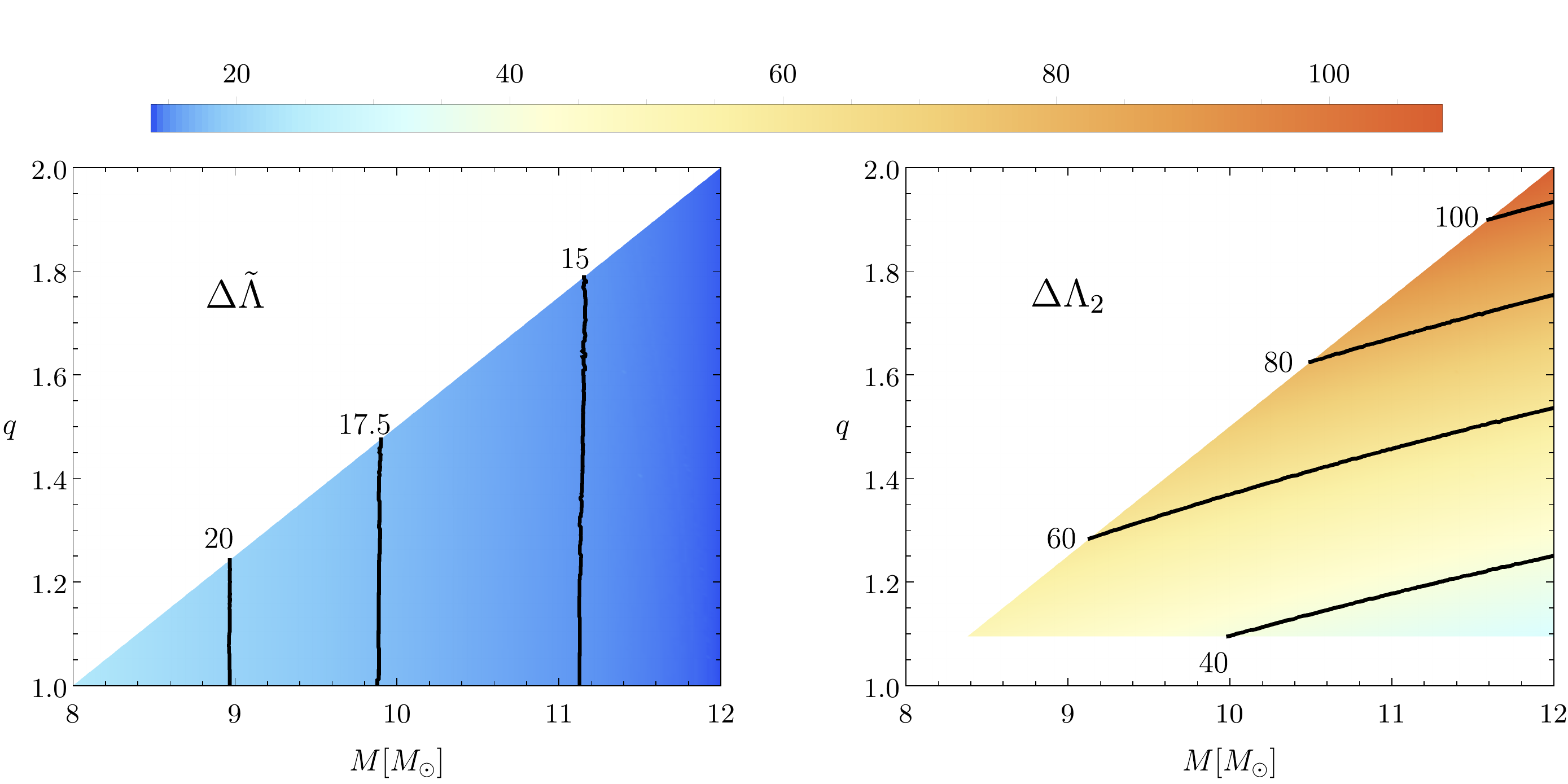}
\caption{Estimated measurement error with Advanced LIGO of (left) the weighted average tidal parameter $\tilde\Lambda$ and (right) the less well-constrained individual tidal parameter $\Lambda_2$ for BBH systems at 400 Mpc. The black lines are contours of constant $\Delta \tilde\Lambda$ and $\Delta \Lambda_2$ in the left and right plots, respectively. }
\label{fig:BBHconstraints}
\end{figure*}

The results for the measurement errors with Advanced LIGO for BBH systems at $400$Mpc are shown in Fig.~\ref{fig:BBHconstraints}, for a starting frequency of $10$Hz. The left panel shows the error in the combination $\tilde \Lambda$ that is directly computed from the Fisher matrix as a function of total mass $M$ and mass ratio ${q=m_1/m_2}$. As discussed above, the ranges of $M$ and $q$ we consider stem from our assumptions on the minimum BH mass and a high merger frequency. The right panel of Fig.~\ref{fig:BBHconstraints} shows the inferred bound on the less well-measured individual deformability in the regime of unequal masses. We omit the region where the objects have nearly equal masses $q\to 1$ because in this regime, the $68\%$ confidence interval $\nu+2\Delta\nu$ exceeds the physical bound $\nu\leq1/4$. Inferring the errors on the parameters of the individual objects requires a more sophisticated analysis~\cite{Cutler:1994ys} than that considered here. The coloring ranges from small errors in the blue shaded regions to large errors in the orange shaded regions; the labeled black lines are representative contours of constant $\Delta\Lambda$. Note that the errors on the individual deformability $\Lambda_2$ are always larger than those on the combination $\tilde \Lambda$. 

We find that the tidal deformability of our fiducial BBH systems can be measured to within $\Delta \Lambda\lesssim 100$ by Advanced LIGO, which indicates that BHs can be readily distinguished from massive BSs. However, even for ideal BBHs---high mass, low mass-ratio binaries---the tidal deformability of each BH can only be measured within  $\Delta \Lambda\gtrsim 15$ by Advanced LIGO. Therefore one cannot distinguish BHs from solitonic BSs using estimates of each bodies' deformability alone. Given these findings, we also estimate the precision with which the tidal deformability could be measured with third-generation instruments. Compared to Advanced LIGO, the measurement errors in the tidal deformability decrease by factors of $\sim13.5$ and $\sim23.5$ with Einstein telescope and Cosmic Explorer, respectively. Thus, the more massive BH in the binary would be marginally distinguishable from a solitonic BS with future GW detectors, as ${\Delta \Lambda_1\leq \Delta \tilde \Lambda \lesssim 1}$. These findings are consistent with the conclusions of Cardoso et al~\cite{Cardoso:2017cfl}, although these authors considered only equal-mass binaries at distances $D=100$Mpc with total masses up to $50M_\odot$.  However, we find that in an unequal-mass BBH case, the less massive body could not be differentiated from a solitonic BS even with third-generation detectors.

\begin{figure*}
\includegraphics[width=0.9\textwidth]{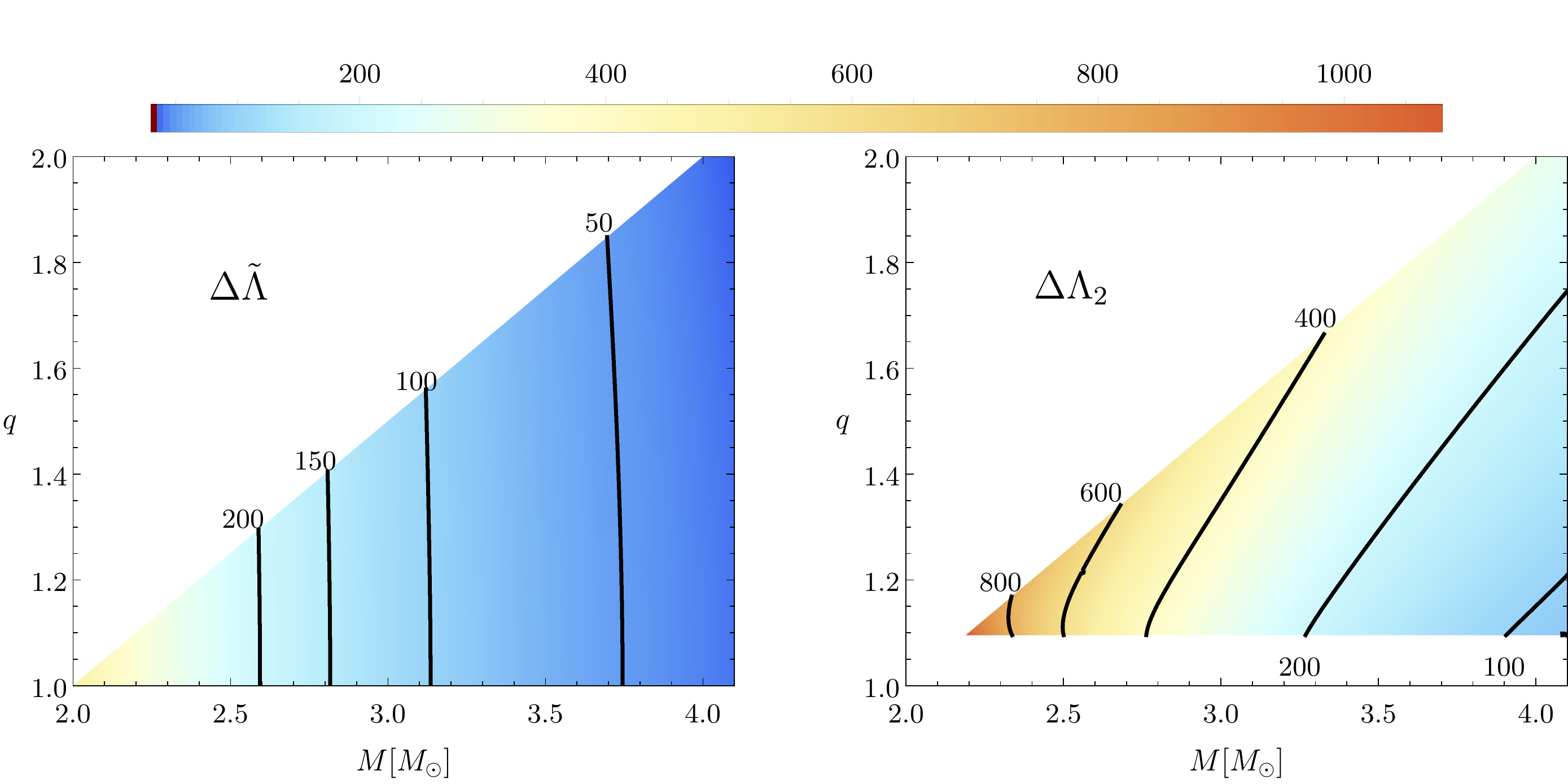}
\caption{Estimated measurement error with Advanced LIGO of (left) the weighted average tidal parameter $\tilde\Lambda$ and (right) the less well-constrained individual tidal parameter $\Lambda_2$ for BNS systems at 200 Mpc with the SLy equation of state. The black lines are contours of constant $\Delta \tilde\Lambda$ and $\Delta \Lambda_2$ in the left and right plots, respectively. }
\label{fig:BNSconstraints}
\end{figure*}

Next, we consider the measurements of a BNS system, shown in Fig.~\ref{fig:BNSconstraints} assuming the SLy EoS. We restrict our analysis to systems with individual masses $1M_\odot\leq m_{\rm NS}\leq m_{\rm max}$, where $m_{\rm max}\approx 2.05 M_\odot$ is the maximum mass for this EoS. Similar to Fig. \ref{fig:BBHconstraints}, the left panel in Fig.~\ref{fig:BNSconstraints} shows the results for the measurement error in the combination $\tilde \Lambda$ directly computed from the Fisher matrix, and the right panel shows the error for the larger of the individual deformabilities. The slight warpage of the contours of constant $\Delta \tilde \Lambda$ compared to those in Fig.~\ref{fig:BBHconstraints}, best visible for the $\Delta \tilde \Lambda=50$ contour, is due to an additional dependence of the merger frequency on $\tilde \Lambda$ for BNSs that is absent for BBHs, and a small difference in the Fisher matrix elements when evaluated for $\tilde \Lambda \neq 0$.
We see that the deformability of NSs of nearly maximal mass in BNS systems can be measured to within $\Delta \Lambda \lesssim 200$, and thus can be distinguished from massive BSs. However, the measurement precision worsens as one decreases the NS mass, rendering lighter NSs indistinguishable from massive BSs using only each bodies deformability alone. In the next subsection, we discuss how combining the measurements of $\Lambda$ for each object in a binary system can improve distinguishability from BSs even when the criteria discussed above are not met.

For completeness, we also computed how well third-generation detectors could measure the tidal deformabilities in BNS systems. As in the BBH case, we find that measurement errors in $\Lambda$ decrease by factors of ${\sim13.5}$ and ${\sim 23.5}$ with the Einstein Telescope and Cosmic Explorer, respectively. However, the conclusions reached above concerning the distinguishability of BHs or NSs and BSs remain unchanged.

\subsection{Distinguishability with a pair of deformability measurements}\label{sec:TwoMeasurements}

In the previous subsection we determined that compact objects whose tidal deformability is much smaller than that of BSs could be distinguished as such with Advanced LIGO, e.g., BHs versus massive BSs. In this subsection, we present a more refined analysis to distinguish compact objects from BSs when the deformabilities of each are of approximately the same size. In particular, we focus on the prospects of distinguishing NSs between one and two solar masses from massive BSs and distinguishing BHs from solitonic BSs. Throughout this section, we only consider the possibility that a single species of BS exists in nature; differentiation between multiple, distinct complex scalar fields goes beyond the scope of this paper. We show that combining the tidal deformability measurements of each body in a binary system can break the degeneracy in the BS model associated with choosing the boson mass $\mu$. Utilizing the mass and deformability measurements of both bodies allows one to distinguish the binary system from a BBS system.

In Figs.~\ref{fig:LambdaMtil} and~\ref{fig:LambdaMtilSolitonic}, the tidal deformability of BSs was given as a function of mass rescaled by the boson mass and self-interaction strength. By simultaneously adjusting these two parameters of the BS model, one can produce stars with the same (unrescaled) mass and deformability. This degeneracy presents a significant obstacle in distinguishing BSs from other compact objects with comparable deformabilities. For example, the boson mass can be tuned for any value of the coupling $\lambda$ ($\sigma_0$) so that the massive (solitonic) BS model admits stars with the exact same mass and tidal deformability as a solar mass NS. However, combining two tidal deformability measurements can break this degeneracy and improve the distinguishability between BSs and BHs or NSs.
As an initial investigation into this type of analysis, we pose the following question: given a measurement $(m_1,\Lambda_1)$ of a compact object in a binary, can the observation $(m_2,\Lambda_2)$ of the companion exclude the possibility that both are BSs? We stress that our analysis is preliminary and that only qualitative conclusions should be drawn from it; a more thorough study goes beyond the scope of this paper. 

\begin{figure}
  \includegraphics[width=\columnwidth]{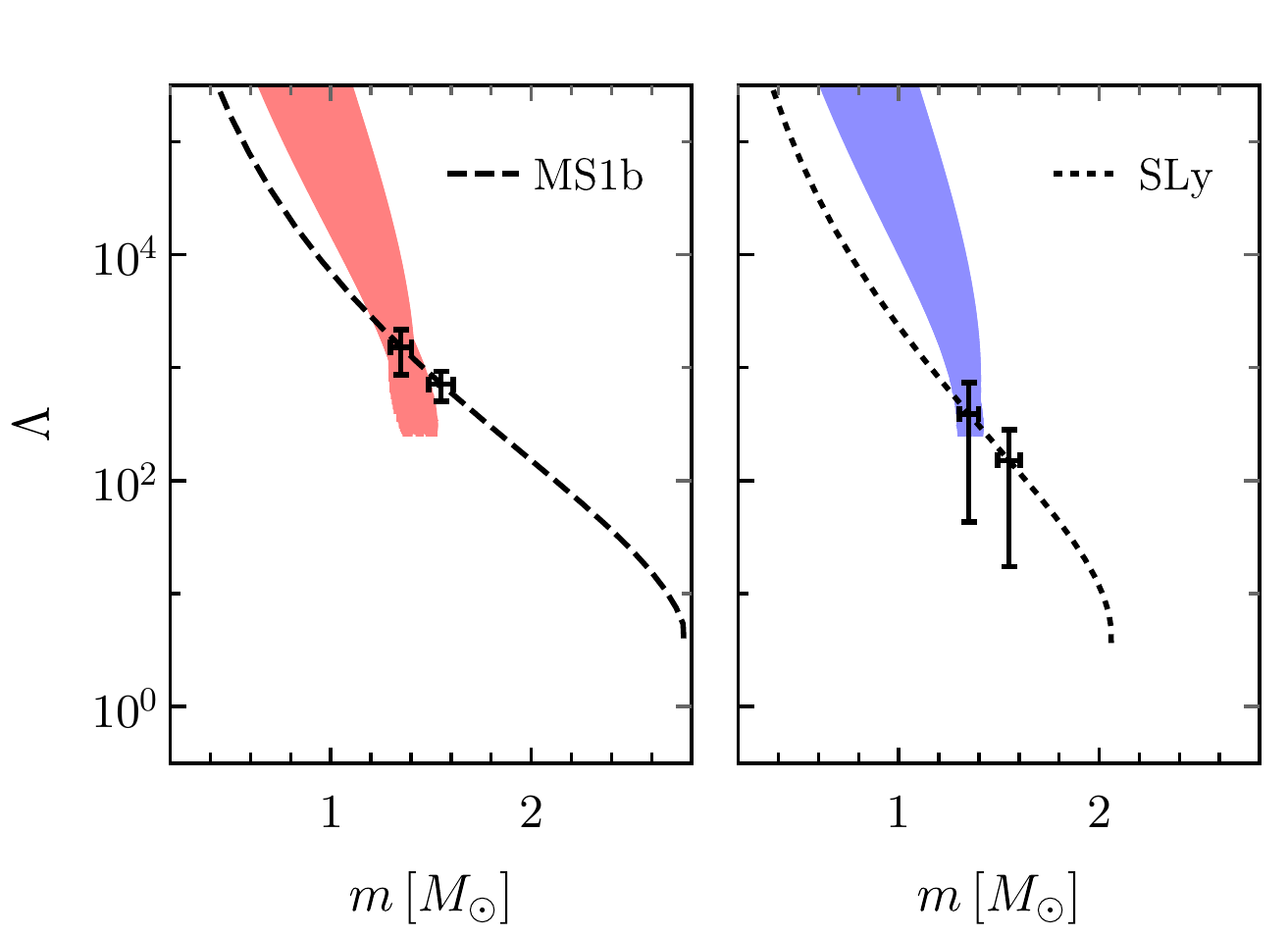}
  \caption{Dimensionless tidal deformability as a function of mass. Black points indicate hypothetical measurements of a ${(1.55+ 1.35) M_\odot}$ binary NS system with the (left) MS1b and (right) SLy EoS; the error bars are estimated for a system observed at 200 Mpc. The shaded regions depict all possible massive BSs (i.e., all possible values of the boson mass $\mu$ and coupling $\lambda$) consistent with the measurement of the smaller compact object. For the MS1b EoS, the tidal deformabilities of the binary are ${\Lambda_{1.55}=714}$ and ${\Lambda_{1.35}=1516}$. For the SLy EoS, the tidal deformabilites are ${\Lambda_{1.55}=150}$ and ${\Lambda_{1.35}=390}$.}\label{fig:LambdaMsunNSNS}
\end{figure}

\begin{figure}
  \includegraphics[width=\columnwidth]{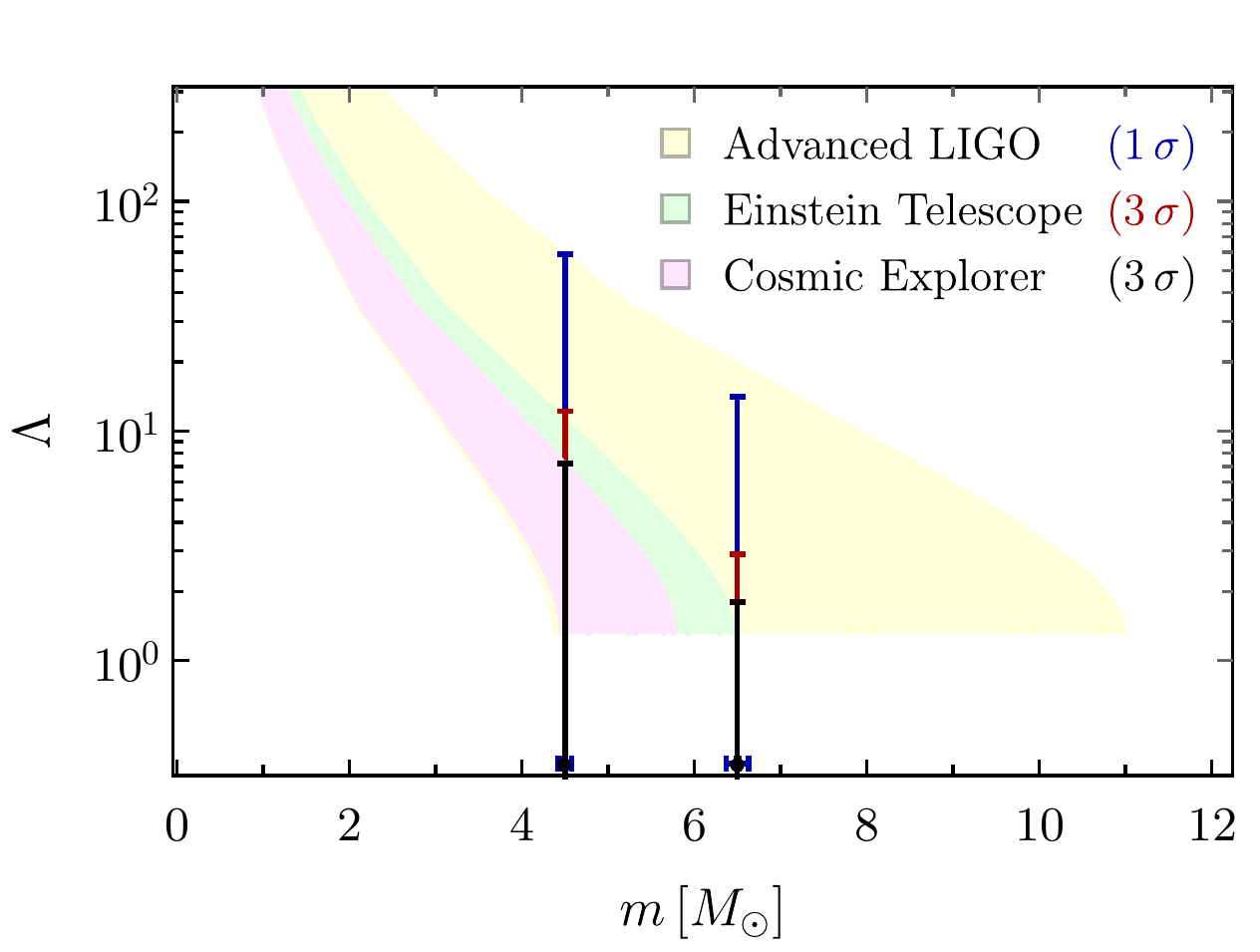}
  \caption{Dimensionless tidal deformability as a function of mass. Hypothetical measurements of a $(6.5+ 4.5) M_\odot$ binary BH system with error bars estimated for a system observed at 400 Mpc by Advanced LIGO, the Einstein Telescope, and Cosmic Explorer are given in blue, red, and black, respectively. The shaded regions depict all possible solitonic BSs with coupling $\sigma_0=0.05\,m_\text{Planck}$ that are consistent with the measurements of the smaller compact object by each detector.}\label{fig:LambdaMsunBHBH}
\end{figure}

From the Fisher matrix estimates for the errors in $({\cal M}, \nu, \tilde \Lambda)$ we obtain bounds on the uncertainty in the measurement $(m_i,\Lambda_i)$ for each body in a binary, which we approximate as being characterized by a bivariate normal distribution with covariance matrix ${\mathbf{\Sigma}=\text{diag}(\Delta m_i,\Delta \Lambda_i)}$. Figure~\ref{fig:LambdaMsunNSNS} depicts such potential measurements by Advanced LIGO of $(m_1,\Lambda_1)$ and $(m_2, \Lambda_2)$, shown in black, for a $({1.55 +1.35}) M_\odot$ BNS system at a distance of 200 Mpc with two representative equations of state for the NSs: the SLy and MS1b models discussed above. The dashed black curves in Figure~\ref{fig:LambdaMsunNSNS} show the $\Lambda(m)$ relation for these fiducial NSs. Figure~\ref{fig:LambdaMsunBHBH} shows the corresponding measurements in a ${6.5 \enDash 4.5 M_\odot}$ BBH measured at 400 Mpc made by Advanced LIGO, Einstein Telescope, and Cosmic Explorer in blue, red, and black, respectively. 

The strategy to determine if the objects could be BSs is the following. Consider first the measurement $(m_2, \Lambda_2)$ of the less massive body. For each point $\mathbf{x}=(m,\Lambda)$ within the $1\sigma$ ellipse, we determine the combinations of theory parameters $(\mu,\lambda)[\mathbf{x}]$ or $(\mu,\sigma_0)[\mathbf{x}]$ that could give rise to such a BS, assuming the massive or solitonic BS model, respectively. As discussed above, in general, $\lambda$ or $\sigma_0$ can take any value by appropriately rescaling $\mu$. Finally, we combine all mass-deformability curves from Figs.~\ref{fig:LambdaMtil} or~\ref{fig:LambdaMtilSolitonic} that pass through the $1\sigma$ ellipise, that is we consider the model parameters ${(\mu,\lambda) \in \bigcup_{\mathbf{x}} (\mu,\lambda)[\mathbf{x}]}$ or ${(\mu,\sigma_0) \in \bigcup_{\mathbf{x}} (\mu,\sigma_0)[\mathbf{x}]}$ for massive and solitonic BS, respectively. These portions of BS parameter space are shown as the shaded regions in Figs.~\ref{fig:LambdaMsunNSNS} and~\ref{fig:LambdaMsunBHBH}. If the tidal deformability measurements $(m_1, \Lambda_1)$ of the more massive body---indicated by the other set of crosses---lie outside of these shaded regions, one can conclude that the measurements are inconsistent with both objects being BSs. 

Figure~\ref{fig:LambdaMsunNSNS} demonstrates that an asymmetric BNS with masses $1.55\enDash1.35M_\odot$ can be distinguished from a BBS with Advanced LIGO by using this type of analysis. When considered individually, either NS measurement shown here would be consistent with a possible massive BS; by combining these measurements we improve our ability to differentiate the binary systems. This type of test can better distinguish BBSs from conventional GW sources than the analysis performed in the previous section because it utilizes measurements of both the mass and tidal deformability rather than just using the deformability alone. However the power of this type of test hinges on the asymmetric mass ratio in the system; with an equal-mass system, this procedure provides no more information than that described in Section~\ref{sec:SingleMeasurement}.

A similar comparison between a BBH with masses $6.5\enDash4.5M_\odot$ and a binary solitonic BS system is illustrated in Fig.~\ref{fig:LambdaMsunBHBH}. For simplicity, the yellow shaded region depicts all possible solitonic BSs for a particular choice of coupling $\sigma_0=0.05\,m_{\rm Planck}$ that are consistent with the measurement of the smaller mass by Advanced LIGO (rather than all possible values of the coupling $\sigma_0$). We see that in contrast to the massive BS case, after fixing the boson mass $\mu$ with the measurement of one body, the measurement of the companion remains within that shaded region. As with the more simplistic analysis performed in  Section~\ref{sec:SingleMeasurement}, we again find that Advanced LIGO will be unable to distinguish solitonic BSs from BHs. 

In the previous section, we showed that third-generation GW detectors will be able to distinguish marginally at least one object in a BBH system from a solitonic BS and thus determine whether a GW signal was generated by a BBS system. Using the analysis introduced in this section, we can now strengthen this conclusion. We repeat the procedure described above for a $6.5\enDash4.5M_\odot$ BBH at 400 Mpc but instead use the $3\sigma$ error estimates in the measurements of the bodies' mass and tidal deformability. In Fig.~\ref{fig:LambdaMsunBHBH}, all possible solitonic BSs consistent with the measurement of the smaller mass are shown in green and pink for Einstein Telescope and Cosmic Explorer, respectively. We see that while the deformability measurements of each BH considered individually are consistent with either being solitonic BSs, they cannot both be BSs. Thus, we can conclude with much greater confidence that third-generation detectors will be able to distinguish BBH systems from binary systems of solitonic BSs.
 
To summarize, the precision expected from Advanced LIGO is potentially sufficient to differentiate between massive BSs and NSs or BHs, particularly in systems with larger mass asymmetry. Advanced LIGO is not sensitive enough to discriminate between solitonic BSs and BHs, but next-generation detectors like the Einstein Telescope or Cosmic Explorer should be able to distinguish between BBS and BBH systems. However, we emphasize again that our conclusions are based on several approximations and further studies are needed to make these precise. We also note that we have deliberately restricted our analysis to the parameter space where waveforms are inspiral-dominated in Advanced LIGO. Tighter constraints on BS parameters are expected for binaries where information can also be extracted from the merger and ringdown portion, provided that waveform models that include this regime are available.

\section{Conclusions} \label{sec:Conclusions}

Gravitational waves can be used to test whether the nature of BHs and NSs is consistent with GR and to search for exotic compact objects outside of the standard astrophysical catalog.  A compact object's structure is imprinted in the GW signal produced by its coalescence with a companion in a binary system. A key target for such tests is the characteristic ringdown signal of the final remnant. However, the small SNR of that part of the GW signal complicates such efforts. Complementary information can be obtained by measuring a small but cumulative signature due to tidal effects in the inspiral that depend on the compact object's structure through its tidal deformability. This quantity may be measurable from the late inspiral and could be used to distinguish BHs or NSs from exotic compact objects. 

In this paper, we computed the tidal deformability $\Lambda$ for two models of BSs: \textit{massive BSs}, characterized by a quartic self-interaction, and \textit{solitonic BSs}, whose scalar self-interaction is designed to produce very compact objects. For the quartic interaction, our results span the entire two-dimensional parameter space of such a model in terms of the mass of the boson and the coupling constant in the potential. For the solitonic case, our results span the portion of interest for BH mimickers. We presented fits to our results for both cases that can be used in future data analysis studies. We find that the deformability of massive BSs is markedly larger than that of BHs and very massive NSs; in particular, we showed that the tidal deformability $\Lambda \gtrsim 280$ irrespective of the boson mass and the strength of the quartic self-interaction. The tidal deformability of solitonic BSs is bounded below by $\Lambda \gtrsim 1.3$. 

To determine whether ground-based GW detectors can distinguish NSs and BHs from BSs, we first computed a lower bound on the expected measurement errors in $\Lambda$ using the Fisher matrix formalism. We considered BBH systems located at 400 Mpc and BNS systems at 200 Mpc with generic mass ratios that merge above 900 Hz. We found that, with Advanced LIGO, BBHs could be distinguished from binary systems composed of massive BSs and that BNSs could be distinguished provided that the NSs were of nearly-maximal mass or of sufficiently different masses (i.e. a high mass ratio binary). We also demonstrated that the prospects for distinguishing solitonic BSs from BHs based only on tidal effects are bleak using current-generation detectors; however, third-generation detectors will be able to discriminate between BBH and BBS systems. We presented two different analyses to determine whether an observed GW was produced by BSs: the first relied on the minimum tidal deformability being larger than that of a NS or BH, while the second combined mass and deformability measurements of each body in a binary system to break degeneracies arising from the (unknown) mass of the fundamental boson field.

Recent work by Cardoso et al.~\cite{Cardoso:2017cfl} also investigated the tidal deformabilities of BSs and the prospects of distinguishing them from BHs and NSs. Despite the topic being similar, the work in this paper is complementary: Cardoso et al.~\cite{Cardoso:2017cfl} performed a broad survey of tidal effects for different classes of exotic objects and BHs in  modified theories of gravity, while our work focuses on an in-depth analysis of BSs. Additionally, these authors computed the deformability of BSs to both axial and polar tidal perturbations with $l=2,3$, whereas our results are restricted to the $l=2$ polar case. The $l=2$ effects are expected to leave the dominant tidal imprint in the GW signal, with the $l=3$ corrections being suppressed by a relative factor of $125 \Lambda_3/(351\Lambda_2)(M \Omega)^{4/3}\sim 4 (M\Omega)^{4/3}$~\cite{Hinderer:2009ca} using the values from Table I of Ref.~\cite{Cardoso:2017cfl}, where $\Omega$ is the orbital frequency of the binary. For reference, $M\Omega\sim 5\times10^{-3}$ for a binary with $M=12M_\odot$ at $900$Hz. 

We also cover several aspects that were not considered in Ref.~\cite{Cardoso:2017cfl}, where the study of BSs was limited to a single example for a particular choice of theory parameters for each potential (quartic and solitonic). Here, we analyzed the entire parameter space of self-interaction strengths for the quartic potential and the regime of interest for BH mimickers in the solitonic case. Furthermore, we developed fitting formulae for immediate use in future data analysis studies aimed at constraining the BS parameters with GW measurements. Cardoso et al.~\cite{Cardoso:2017cfl} also discussed prospective constraints obtained from the Fisher matrix formalism for a range of future detectors, including the space-based detector LISA that we did not consider here. However, their analysis was limited to equal-mass systems, to bounds on $\tilde \Lambda$, and to the specific examples within each BS models. We went beyond this study by delineating a strategy for obtaining constraints on the BS parameter space from a pair of measurements and considering binaries with generic mass ratio. We also restricted our results to the regime where the signals are dominated by the inspiral. Although this choice significantly reduces the parameter space of masses surveyed compared to Cardoso et al., we imposed this restriction because full waveforms that include the late inspiral, merger and ringdown are not currently available. Another difference is that we took BBH or BNS signals to be the ``true" signals around which the errors were computed and used results from NR for the merger frequency to terminate the inspiral signals, whereas the authors of Ref.~\cite{Cardoso:2017cfl} chose BS signals for this purpose and terminated them at the Schwarzschild ISCO. 

The purpose of this paper was to compute the tidal properties of BSs that could mimic BHs and NSs for GW detectors and to estimate the prospects of discriminating between such objects with these properties. Our analysis hinged on a number of simplifying assumptions. For example, the Fisher matrix approximation that we employed only yields lower bounds on estimates of statistical uncertainty. Additionally, we considered only a restricted set of waveform parameters, whereas including spins could also worsen the expected measurement accuracy. On the other hand, improved measurement precision is expected if one uses full inspiral-merger-ringdown waveforms or if one combines results from multiple GW events. Our conclusions should be revisited using Bayesian data analysis tools and more sophisticated waveform models, such as the EOB model. Tidal effects are a robust feature for any object, meaning that the only change needed in existing tidal waveform models is to insert the appropriate value of the tidal deformability parameter for the object under consideration. However, the merger and ringdown signals are more difficult to predict, and further developments and NR simulations are needed to model them for BSs or other exotic objects.

\acknowledgments
N.S. acknowledges support from NSF Grant No. PHY-1208881. We thank Ben Lackey for useful discussions. We are grateful to Vitor Cardoso and Paolo Pani for helpful comments on this manuscript.

\bibliography{./inspire}

\end{document}